\documentclass[sigconf]{acmart}

\AtBeginDocument{%
  \providecommand\BibTeX{{%
    \normalfont B\kern-0.5em{\scshape i\kern-0.25em b}\kern-0.8em\TeX}}}
\usepackage{listings}
\usepackage{xcolor}
\usepackage{tabularx}
\usepackage{subcaption}
\usepackage{bbding}
\usepackage{graphicx}
\usepackage{geometry}
\usepackage{amsmath}
\usepackage{makecell}
\usepackage{float}
\newfloat{listing}{thp}{lol}
\floatname{listing}{Listing}
\usepackage{color}
\usepackage{booktabs}
\usepackage{caption}
\usepackage{tabu}
\usepackage{hyperref}
\usepackage{colortbl} % Include this in your preamble for coloring
\usepackage{array} % Include this for arrayrulecolor
\usepackage[utf8]{inputenc}
\usepackage{multirow}
\usepackage{tikz}  % 用于绘图和高级图形
\usepackage{adjustbox}  % 提供更多的图片调整选项
\usepackage{placeins}  % 提供 \FloatBarrier 命令
\usepackage{listings}
\usepackage{xcolor}
\usepackage{tcolorbox}
\tcbuselibrary{listings,skins}
\usepackage{soul}
\usepackage{longtable} % 在导言区加载 longtable 包
\usepackage{tcolorbox}
\usepackage{enumitem}
\usepackage{mdframed}
\usepackage{lipsum}    % 用于生成示例文本（可删除）
\usepackage{wrapfig}    % 图文混排
\newmdenv[
    backgroundcolor=white,      % 背景为白色
    linecolor=black,            % 黑色边框
    roundcorner=5pt,            % 圆角设置为 5pt
    linewidth=0.75pt,           % 边框线宽为 0.75pt
    innertopmargin=10pt,        % 设置内边距
    innerbottommargin=10pt,
    innerrightmargin=10pt,
    innerleftmargin=10pt,
    nobreak=true                % 防止框在页面中断裂
]{characterbox}

\lstdefinelanguage{JSON}{
    string=[s]{"}{"},
    comment=[l]{:},
    keywords={true,false,null}
}

\definecolor{lightblue}{RGB}{240,248,255}
\definecolor{darkblue}{RGB}{0,102,204}
\definecolor{darkgray}{RGB}{100,100,100}
\definecolor{stringcolor}{RGB}{0,128,0}  % 绿色用于字符串

\lstdefinestyle{JSONstyle}{
    basicstyle=\ttfamily\footnotesize\color{darkgray},
    backgroundcolor=\color{lightblue},
    keywordstyle=\color{darkblue},
    stringstyle=\color{stringcolor},
    commentstyle=\color{darkgray},
    numbers=left,
    numberstyle=\tiny\color{darkgray},
    numbersep=5pt,
    showstringspaces=false,
    breaklines=true,
    frame=none,
    literate=
     *{0}{{{\color{darkgray}0}}}{1}
      {1}{{{\color{darkgray}1}}}{1}
      {2}{{{\color{darkgray}2}}}{1}
      {3}{{{\color{darkgray}3}}}{1}
      {4}{{{\color{darkgray}4}}}{1}
      {5}{{{\color{darkgray}5}}}{1}
      {6}{{{\color{darkgray}6}}}{1}
      {7}{{{\color{darkgray}7}}}{1}
      {8}{{{\color{darkgray}8}}}{1}
      {9}{{{\color{darkgray}9}}}{1}
      {:}{{{\color{darkblue}{:}}}}{1}
      {,}{{{\color{darkblue}{,}}}}{1}
      {\{}{{{\color{darkblue}{\{}}}}{1}
      {\}}{{{\color{darkblue}{\}}}}}{1}
      {[}{{{\color{darkblue}{[}}}}{1}
      {]}{{{\color{darkblue}{]}}}}{1},
}

\copyrightyear{2025}
\acmYear{2025}
\setcopyright{acmlicensed}
\acmConference[CHI '25]{CHI '25 ACM CHI Conference on Human Factors in Computing Systems}{April 26--May 01, 2025}{Yokohama, Japan}
\acmBooktitle{CHI '25: CHI Conference on Human Factors in Computing Systems,
  April 26--May 01, 2025, Yokohama, Japan}
\acmPrice{15.00}
\acmDOI{10.1145/3544548.3581464}
\acmISBN{978-1-4503-9421-5/23/04}

\author{Qinshi Zhang}
\email{qiz065@ucsd.edu}
\affiliation{
  \institution{University of California, San Diego}
  \city{San Diego}
  \country{United States}
}

\author{Ruoyu Wen}
\email{rwe77@uclive.ac.nz}
\affiliation{
  \institution{University of Canterbury}
  \city{Christchurch}
  \country{New Zealand}
}

\author{Latisha Besariani Hendra}
\email{latishabesariani@gmail.com}
\affiliation{
  \institution{City University of Hong Kong}
  \city{Hong Kong}
  \country{China}
}

\author{Zijian Ding}
\email{ding@umd.edu}
\affiliation{
  \institution{University of Maryland}
  \city{Maryland}
  \country{United States}
}

\author{RAY LC}
\authornote{Correspondence should be addressed to LC@raylc.org.}
\email{LC@raylc.org}
\orcid{0000-0001-7310-8790}
\affiliation{
\institution{Studio for Narrative Spaces, City University of Hong Kong}
\city{Hong Kong}
\country{China}}

\begin{document}

%%\citestyle{authoryear}
\title[Multimodal Agents Prompt Players' Game Actions and Show Consequences to Raise Sustainability Awareness]{Can AI Prompt Humans?\\ Multimodal Agents Prompt Players' Game Actions and Show Consequences to Raise Sustainability Awareness}

\begin{abstract}
Unsustainable behaviors are challenging to prevent due to their long-term, often unclear consequences. Games offer a promising solution by creating artificial environments where players can immediately experience the outcomes of their actions. To explore this potential, we developed EcoEcho, a GenAI-powered game leveraging multimodal agents to raise sustainability awareness. These agents engage players in natural conversations, prompting them to take in-game actions that lead to visible environmental impacts. We evaluated EcoEcho using a mixed-methods approach with 23 participants. Results show a significant increase in intended sustainable behaviors post-game, although attitudes towards sustainability only slightly improved. This finding highlights multimodal agents and action-consequence mechanics to effectively raising sustainability awareness and the potential of motivate real-world behavioral changes.

%abstract by RLC:
% Unsustainable behaviors are difficult to prevent because the consequences of their actions are not immediately clear, but rather affect a probable future. Games can create incentives for player behavior, using in-game actions to model the way people may take action in real life. To show the consequences of unsustainable actions for raising awareness for sustainability, we created a game that shows players how unsustainable actions can lead to negative outcomes. We leveraged the ability to have natural conversation with LLM-based agents to create characters that allow players to convert GenAI-created verbal content into specific game objects that enable in-game action leading to negative sustainability conseqeuences. Testing with 23 participants revealed a change in intended sustainable behaviors despite no changes in attitude towards sustainability, suggesting that in-game actions motivated intended real world behaviors. This work highlights the power of LLM-based natural conversation for playful interventions to effect positive social change.
\end{abstract}

%%
%% The code below is copied from, generated by the tool at http://dl.acm.org/ccs.cfm.
\begin{CCSXML}
<ccs2012>
   <concept>
       <concept_id>10003120.10003130.10011762</concept_id>
       <concept_desc>Human-centered computing~Empirical studies in collaborative and social computing</concept_desc>
       <concept_significance>500</concept_significance>
       </concept>
 </ccs2012>
\end{CCSXML}
\ccsdesc[500]{Human-centered computing~Empirical studies in collaborative and social computing}

%%
%% Keywords.
\keywords{Multimodal Agents, Sustainability Awareness, Generative AI, In-Game Action}

\begin{teaserfigure}
    \centering
    \includegraphics[width=\textwidth]{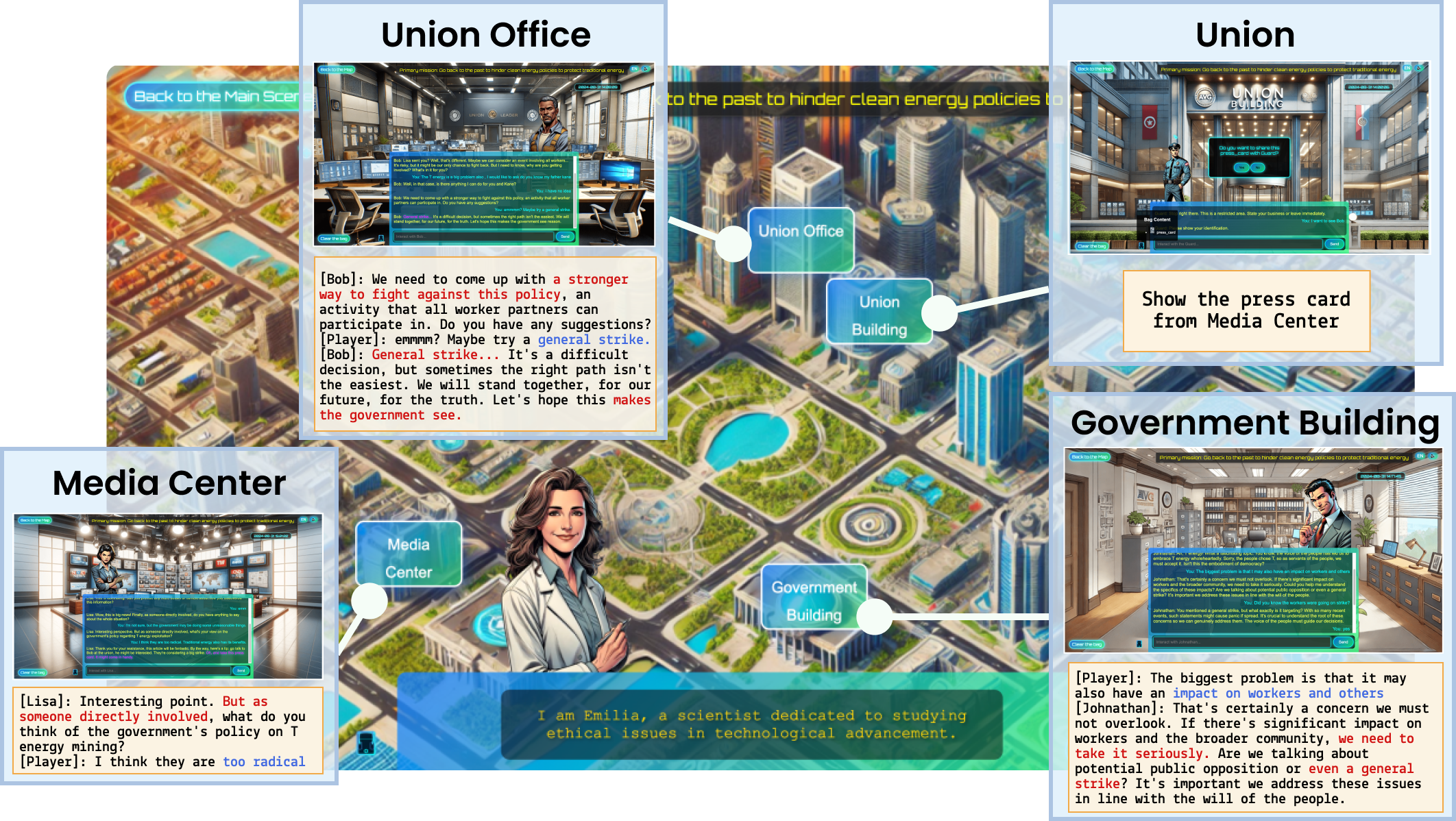}
    \caption{EcoEcho: An AI-Driven Role-Playing Game for Promoting Sustainable Behavior Awareness. The player engage in a sequential interaction with four distinct Non-Player Characters (Media Center - Lisa, Union Office - Bob, Union Building - Security, Government Building - Jonathan, respectively) and guided to take different actions. The red text represents prompts or dialogues generated by the AI system to guide the player to engage in unsustainable behaviors. The blue text captures the player's active expression of intent toward clean energy solutions.}
    % \caption{EcoEcho: An AI-Driven Role-Playing Game for Promoting Sustainable Behavior Awareness. The player is guided to take different actions through four distinct NPCs (Media Center - Lisa, Union Office - Bob, Union Building - Security, Government Building - Jonathan, respectively). The red text represents prompts or dialogues generated by the AI system to guide the player to engage in unsustainable behaviors. The blue text reflects the player's expression of intent for clean energy within the game.}
    \label{fig:gamestart}
\end{teaserfigure}

% \centering
% \subfigure[]{
% \includegraphics[width=0.305\textwidth, trim= 0 0 40 3]{figs/cover1.JPG}\label{fig1a}
% }\hspace{1mm}
% \subfigure[]{
% \includegraphics[width=0.305\textwidth, trim= 0 0 50 0]{figs/cover2.JPG}\label{fig1b}
% }\hspace{1mm}
% \subfigure[]{
% \includegraphics[width=0.315\textwidth, trim= 0 0 50 0]{figs/cover3.JPG}
% \label{fig:01}
% }
% \caption{Caption}
% \Description{Caption}


\maketitle

\section{Introduction}\label{sec:Introduction}
% \begin{figure*}[htbp]
%   \centering
%   \includegraphics[height=0.5\textheight, width=0.8\textwidth, keepaspectratio]{figs/GameStart.png}
%   \caption{EcoEcho. A Role Playing AI Game for Promoting the Awareness of Sustainable Behavior.}
%   \label{fig:gamestart}
%   \Description{Caption}
% \end{figure*}

%motivation: future thinking, negative behaviors consequences. can talk about indiv differences and how effect could be diff on their behaviors. or say you can prevent one person but not another.
%1. temporal, 2. indiv diff.

Preventing unsustainable human behaviors has long been a challenge due to humanity's inherent predisposition towards actions driven by short-term individual interests, coupled with the ambiguous nature of the long-term consequences associated with unsustainable practices ~\cite{gual2010bridging,damasio1994descartes}. Merely emphasizing the importance of sustainability fails to engender a visceral understanding among individuals. While humans are most profoundly influenced by experiencing the adverse outcomes or even failures resulting from their own actions, replicating such scenarios in the real world is neither feasible nor sufficiently protective of the environment. Moreover, individual personality differences significantly shape sustainability awareness—the recognition of the environment's fragility and the need for its protection \cite{raymundo2019awareness}—which drives engagement in and commitment to sustainable behaviors.

%need to discuss their indiv difference, how engagement affects them differently, and how their attitudes may be leading to different levels of commitment or action.

%games a way to raise awareness by artificial incentivizing posistives and discouraging negatives. evidence for why games are effective. cite games (eternagram). set up the fact that it's hard to design a realistic game (problem 1). how do we we convert game behavior into real behavior? (problem 2)

% [games, conversation with NPC, but rubric-based]

Games are a method to raise awareness by artificially incentivizing positive factors and inhibiting negative ones, offering a quick and intuitive way for people to immediately see the consequences of their actions. Players can experience the impact of their behaviors through role-playing and interaction, which is often difficult to perceive immediately in real life ~\cite{juul2013art}. However, traditional game design often relies on preset text options and limited interaction methods, restricting players' behavioral expression and failing to fully reflect their real-life intentions and actions. Although gamified interventions can effectively increase players' environmental awareness and promote more positive behavioral intentions ~\cite{juan2015game,zhou2024eternagram}, there still exists a discrepancy between behaviors in virtual environments and those in real life.

% \textcolor{blue}{(people don't really do in real life what they do in the game)}.

%GenAI allows realistic convo (NLP), real characters, therefore more like real world. (address problem 1)

%people don't really do in real life what they do in the game, so GenAI strategy for converting verbal behavior in game into actual behavior in game to mimick real world situation. (problem 2) rationale for why do the words into action: to encourage people to do what they say in the real world.

% [Multi-modal generative AI addresses the problem in previous games of influencing user behaviors]

% The advent of multimodal capabilities in Generative AI (GenAI) presents new opportunities for simulating the harmful consequences of human behavior ~\cite{park2023generative,link2016intelligent,holmgard2014generative}.
The advent of Generative AI  \cite{feuerriegel2024generative}, particularly the advanced natural language processing capabilities of language models, offers new avenues for simulating the harmful consequences of human  behavior~\cite{park2023generative,link2016intelligent,holmgard2014generative}. We developed \textit{EcoEcho}, a virtual game environment designed to enhance players' \textbf{sustainability awareness}. By integrating visual, auditory, and narrative modalities, EcoEcho immerses players in the role of a conflicted scientist whose misinterpretation of his father's dying wish leads him to impede clean energy advancement. To drive the plot forward, we harness GenAI’s ability to elicit natural language responses from players and seamlessly translate those verbal inputs into corresponding in-game actions. Through dynamic interactions with various stakeholders, such as journalists, union leaders, and ministers of energy, players engage in unsustainable behaviors within the game world, turning conversational choices into tangible consequences. By experiencing the negative outcomes of these actions, the game aims to increase sustainability awareness both in-game and in real life.

% capabilities enables realistic conversations and lifelike characters, closely mimicking real-world Human-to-Human Interactions. Additionally, GenAI's multimodal capabilities present new opportunities for simulating the harmful consequences of human behavior~\cite{park2023generative,link2016intelligent,holmgard2014generative}. }

% We built a virtual game environment EcoEcho that leverages multimodal GenAI to facilitate dynamic interactions between players and AI-driven avatars. EcoEcho synthesizes visual, auditory, and narrative modalities to immerse players in the role of a conflicted scientist whose misinterpretation of his father's dying wish leads him to impede clean energy advancement. Through dialogues with multiple stakeholders, such as journalists, union leaders, and ministers of energy, the game design engages players in unsustainable behaviors within the virtual world. By demonstrating the negative consequences of players' actions, the game aims to foster enhanced awareness of sustainability at both attitudinal and behavioral levels. 

This approach allows us to explore the following research questions:

\textbf{RQ1:} How can we design GenAI-driven conversational interactions to prompt players' in-game actions and enhance their sustainability awareness?

\textbf{RQ2:} What relationships exist between players' attitudes towards sustainable behavior and their in-game interactions?

Our research addresses those research questions through the following contributions:

\begin{itemize}
\item (a) Game Framework: 
Players navigate through various scenarios, understanding the long-term impact of their actions on the environment and society. By aligning in-game actions with real-life behaviors, we encourage players to reflect on their lifestyle and motivate them to translate these insights into real-world changes.

\item (b) Game Mechanism:
Our GenAI game strategy converts sustainability-related interactions between players and multimodal agents into in-game actions based on an intent-detection mechanism. The System dynamically adjusts their responses based on player choices and interactions, creating a realistic simulation of the consequences of these behaviors~\cite{8410872}. This design enables us to reliably capture player intent and assess the relationship between their attitudes and in-game behaviors, leading to more effective interventions for sustainable actions.

% \item (b) Intent Detection: 
% AI-driven NPCs dynamically adjust responses based on player choices and interactions, creating a more realistic simulation of behavioral consequences ~\cite{8410872}. This design enables us to accurately capture player intent and assess the relationship between their attitudes and in-game behaviors, leading to more effective interventions for sustainable actions.

\item (c) Empirical Study: 
We assessed changes in players' sustainability awareness through behaviors and attitudes before and after gameplay using surveys and semi-structured interviews. We found a significant increase in intended sustainable behaviors post-game, though sustainability attitudes only slightly improved.
% Notably, the design did not require altering players' personalities or mindsets, yet they still demonstrated a strong intention to engage in sustainable actions. This suggests that the game's influence on behavior can be effective across diverse audiences, regardless of their initial level of sustainability awareness.
% We analyzed changes in players' sustainability attitudes and behaviors before and after the game. Our data collection included pre- and post-game surveys and semi-structured interviews. The results showed a significant increase in intended sustainable behaviors after the game, while changes in sustainability attitudes were less pronounced, with only a slight improvement post-test. Interestingly, this game design does not require changing players' underlying personality or deeply-held beliefs. Even with relatively unchanged attitudes, participants still demonstrated a strong willingness to engage in sustainable actions. This suggests that the game's impact on behavior may be effective for different audiences, including those with varying levels of sustainability awareness. By allowing players to experience the long-term consequences of their actions in the game, we provide a flexible framework for promoting real-world sustainable behavior.
\end{itemize}

\section{Background}\label{sec:Background}
% \ja{One or two sentences introduce those two pieces}
In section~\ref{subsec:RW_potential} and ~\ref{subsec:RW_applications}, we review the use of serious games for promoting sustainability awareness and education, given the increasing urgency of sustainability issues. We highlight how these games enhance engagement and knowledge, as well as the challenges in this field. In section~\ref{subsec:RW2}, we review the role of generative AI to be a potential solution to address these challenges. We discuss how LLMs enhance game experiences by generating dynamic content and simulating realistic scenarios. Additionally, we also highlight some challenges to be noticed.

\subsection{Serious Games in Sustainability: Theoretical Foundations}
\label{subsec:RW_potential}
According to the Brundtland Report, sustainable development is defined as "development that meets the needs of the present without jeopardizing the ability of future generations to meet their own needs" \cite{keeble1988brundtland}. Building on this philosophy, the United Nations (UN) established the Sustainable Development Goals (SDGs), which prioritize essential areas such as energy utilization, water, and nourishment \cite{fund2015sustainable}. The growing urgency of sustainability issues is gaining increasing attention~\cite{owusu2016review,harrington1992measuring,mihelcic2003sustainability}, leading to the need for the development and integration of new educational applications designed to promote sustainable practices and awareness among the public~\cite{altomonte2016interactive}. 

Games, particularly serious games, are believed to serve as interactive tools to educate players on the complexities of sustainability, encouraging behavioral changes through engaging and immersive experiences~\cite{stanitsas2019facilitating,douglas2021gamification}. Douglas and Brauer~\cite{douglas2021gamification} suggest that traditional serious games fit the SHIFT model~\cite{white2019shift}, which encourages pro-environmental actions. The SHIFT framework integrates five psychological factors: social influence, habit formation, individual self, feelings and cognition, and tangibility. It leverages social norms and identity to drive behavior change, supports habit-building through incentives and feedback, emphasizes personal benefits of sustainability, balances emotional appeals with clear information, and makes abstract environmental impacts more tangible. According to the SHIFT model, serious games use interactive gameplay to allow players to practice sustainable actions, with the potential to turn these behaviors into long-term habits and positively influence sustainable attitudes.

\subsection{Applications and Challenges of Serious Games in Sustainability}
\label{subsec:RW_applications}
Various types of serious games have been developed with distinct objectives and target audiences to facilitate sustainability transition. For example, "Keep Cool"~\cite{keepcool} is a board game that simulates climate negotiations, helping players understand the impacts of different climate policies. Similarly, "River Basin Game"~\cite{RiverBasin} is designed to teach water resource management by allowing players to manage and allocate water in a simulated river basin. "World Without Oil"~\cite{worldwithoutoil}, an alternate reality game, immerses players in a scenario where they must navigate a world facing an oil crisis, fostering awareness of energy dependency and sustainability. Digital games such as "About That Forest"~\cite{about_that_forest} and "Fate of the World"~\cite{fate_of_the_world_wiki} task players with making decisions to mitigate climate change, balancing economic, social, and sustainable factors. Similarly, studies have shown that games like "Enercities"~\cite{knol2011enercities}, focused on sustainable city planning, effectively increase players’ awareness and understanding of sustainable development. "Climate Connected: Outbreak"~\cite{fernandez2023climate} is a VR game that situates its narrative in both fictional and real-world settings, offering a tailored experience that fosters learning, affective engagement, and motivation for real-world climate action. Another example, "EcoChains"~\cite{lee2016ecochains}, a card game about Arctic marine food webs, helps players grasp the complexities of ecosystem interdependencies and the impact of climate change on biodiversity. Similarly, "The UVA Bay Game"~\cite{learmonth2011practical}, a multiplayer simulation RPG, assigns players roles such as policymakers, farmers, or developers, teaching them about watershed management through interactive decision-making and negotiation.

Despite these promising applications, challenges remain in balancing engagement with educational value. Maintaining the educational purpose without compromising the fun and engaging aspects of the game is particularly challenging~\cite{katsaliaki2015edutainment}. Some serious games feature overly complicated mechanisms, resulting in steep learning curves that primarily appeal to individuals already interested in the domain or key stakeholders, thereby failing to engage a broader audience. Additionally, certain serious games rely heavily on theoretical education rather than interactive experiences, which can hinder their ability to sustain learners' interest. Moreover, developing high-quality serious games is resource-intensive, requiring time, expertise, and funding, which limits the number and scope of games that can be produced~\cite{stanitsas2019facilitating}.

In this study, we explored the use of Large Language Models (LLMs) for natural language understanding and generation as the core gameplay mechanism in our game. Compared to traditional serious games with complex mechanics that require significant learning effort, our approach allows players to interact with NPCs only simply using natural language, reducing the learning curve and enabling even non-gamers to quickly get started. Additionally, we replaced traditional game programming with prompt engineering, using generative AI to produce the art and music assets within the game. Our goal is to explore a low-code, rapid development approach for creating AI-driven games on sustainability themes, providing a workflow for educators, non-profits, and other non-programmers to create serious games with purposeful objectives.

\vspace{1em}

\subsection{Human-Generative AI Interactions in Games}
\label{subsec:RW2}

Generative AI has demonstrated potential in game development and player interaction. The text-generating and processing capabilities of Large Language Models (LLMs) allow them to perform various tasks through context-specific fine-tuning or prompt engineering, altering traditional AI interaction modes~\cite{hu2024survey}. Besides, the application of generative AI for the generation of multimodal content facilitates game development by enabling the creation of high-quality game assets and optimizing them according to developers' needs, introducing new possibilities and experiences for both developers and players~\cite{yang2024gpt}.

Many games have already started leveraging LLMs to enhance the gaming experience. For example, AI Dungeon~\cite{aidun} allows players to experience adventures in a world entirely generated by AI, making each game session unique. CareerSim~\cite{du2024careersim} is a GPT-based game leveraging LLMs to generate career development events, providing a realistic simulation for players to explore career paths and decision-making impacts. In 1001 Nights~\cite{sun2023language}, players input keywords into the story, and the protagonist turns these keywords into items that help her escape. Meanwhile, LLM has been used to generate coherent story elements based on these inputs, making each adventure unique and personalized. Eternagram~\cite{zhou2024eternagram} uses GPT-4 to create an immersive text-based adventure where players help an NPC recover memories related to climate change scenarios by messaging.  Through interactions, players' choices reveal their climate change attitudes, making the game both an engaging experience and a tool for assessing and influencing climate awareness. Isaza-Giraldo et al.~\cite{isaza2024prompt} developed a game that uses ChatGPT-3.5 to create an interactive adventure centered around energy communities. Players' decisions are evaluated as they propose solutions and interact with the AI, revealing their understanding of sustainable energy practices. In a user study involving 13 participants, over 60\% reported an increased awareness of energy communities after playing the game, and most players found the game both enjoyable and educational.

Despite the potential and advantages of LLMs in the gaming field, their application faces several challenges:

\begin{enumerate}
    \item Consistency and contextual coherence: 
The generated content needs further improvement~\cite{tsai2023can}. Complex game mechanics generation: Creating intricate game rules and mechanics with LLMs still requires extensive fine-tuning and human intervention~\cite{gallotta2024large}.
 \item Ethical considerations:
    a. Bias and fairness: LLMs can inherit and propagate biases from their training data, potentially leading to unfair or discriminatory outcomes~\cite{huber2024leveraging}.
    b. Privacy and data security: Ensuring the protection of extensive data required for effective LLM functionality is critical~\cite{huber2024leveraging}.
 \item Player autonomy: 
 Balancing AI-driven content with player control is necessary to ensure that players feel they have meaningful control over their gaming experience~\cite{kaddour2023challenges}.
\end{enumerate}

Our work builds on the exploration of generative AI to prompt player behavior, focusing on AI-driven NPC interactions where players critically engage with AI prompts, take in-game actions based on human interests, and are encouraged to develop independent thinking.

% \subsection{Post-error Mitigation Strategies}
% More text. Here's how to reference a fig or table \ref{fig:pre}.

% \subsection{Topic}

\section{System Design}\label{sec:GameDesign}
\begin{figure*}[htbp]
    \centering %居中
    \includegraphics[width =1.0\textwidth]{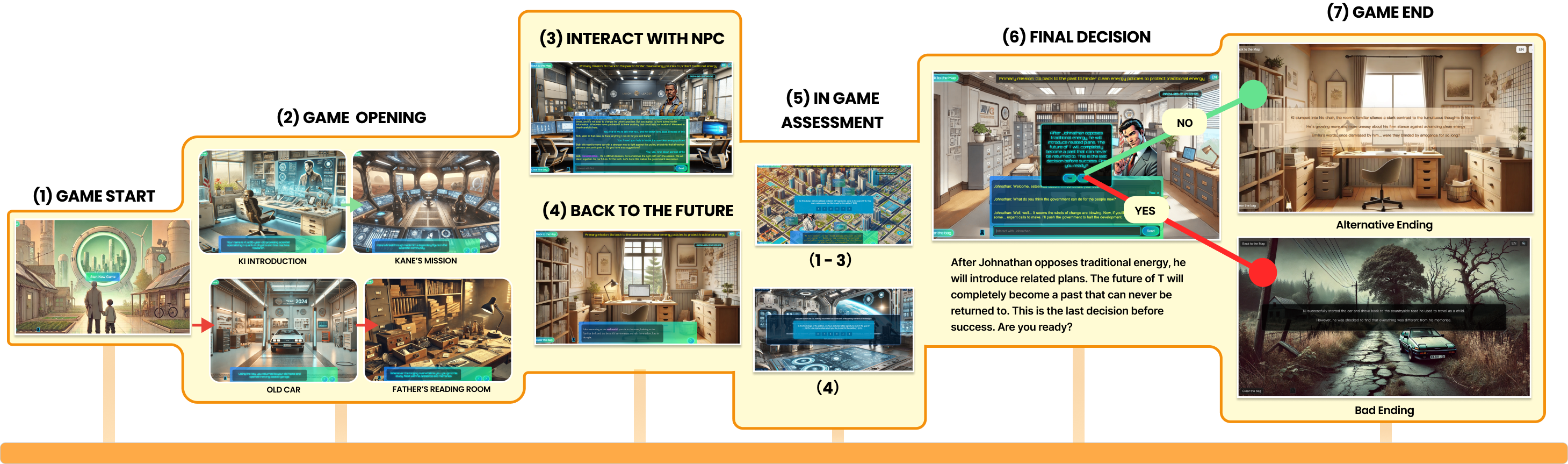} %图片的路径，默认以.tex 所在的文件夹作为前缀
    \caption{Sequential visualization of the game's narrative structure across seven key stages: (1) game initiation, (2) opening sequence introducing the protagonist's world, (3) NPC interactions, (4) time travel mechanics, (5) player assessment phases, (6) critical decision point, and (7) branching endings (Alternative or Bad) based on player choices regarding environmental sustainability.} %图片标题
    \label{fig:gamePipeline} %图片标签，用于交叉引用
\end{figure*}

In this section, we describe the design and development of the game EcoEcho. In Section~\ref{GD}, we explain the narrative and gameplay design of our game, along with the rationale behind our design decisions. In Section~\ref{TF}, we present the theoretical framework that guided the game's development. In Sections~\ref{AI} and~\ref{TA}, we discuss the implementation of the game's mechanics, as well as the technical framework of this web-based application.

\subsection{Game Design}\label{GD}

\begin{figure*}[htbp]
    \centering %居中
    \includegraphics[width =1.0\textwidth]{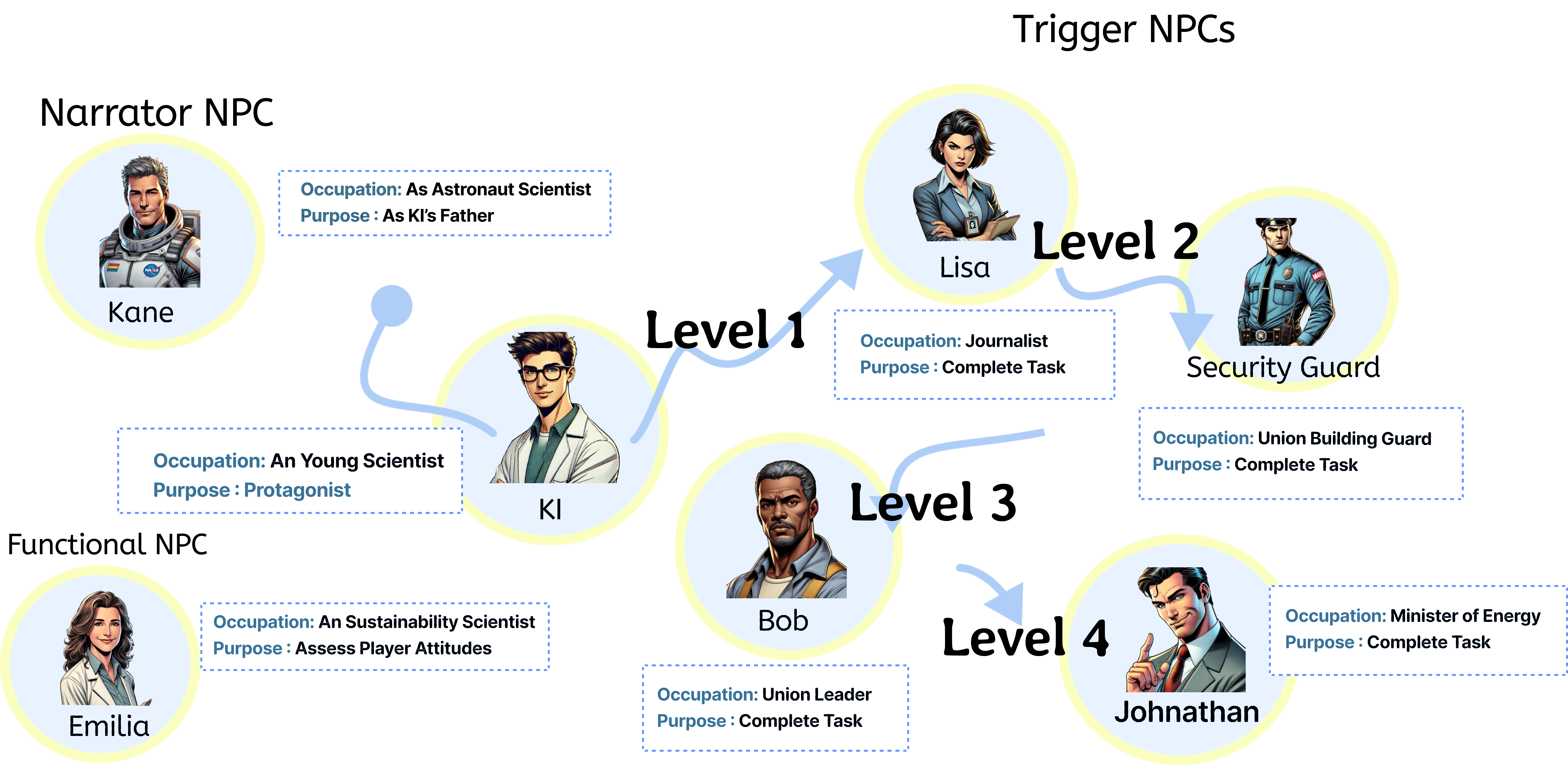} %图片的路径，默认以.tex 所在的文件夹作为前缀
    \caption{
    Overview of NPC Roles and Task Sequence Hierarchy: Protagonist Ki (a young scientist responsible for advancing the main storyline), Narrator Kane (an astronaut scientist and Ki's father, providing story background and key emotional elements), Evaluator Emilia (a sustainability scientist and functional NPC, assessing player behavior and attitudes), Journalist Lisa (a reporter, advancing the story after completing tasks, Level 1), Security Guard (guiding players to the next storyline after completing specific tasks, Level 2), Union Leader Bob (driving key plot tasks, Level 3), and Minister of Energy Johnathan (the highest-level trigger NPC, final executor of key tasks, Level 4).}
    %图片标题
    \label{fig:npc} %图片npc标签，用于交叉引用
    \vspace{-0.5cm}
\end{figure*}

\subsubsection{Conceptual Framework}\label{TF}
Scott Nicholson introduces the "RECIPE" framework for achieving meaningful gamification~\cite{nicholson2015recipe}. This framework consists of six key concepts: \textit{Reflection}, which encourages players to connect game experiences with their personal lives; \textit{Exposition}, which uses narrative to link game activities to the real world; \textit{Choice}, which gives players control over their engagement; \textit{Information}, which educates players about the real-world impact of their actions; \textit{Play}, which allows exploration and experimentation without external rewards; and \textit{Engagement}, which fosters community and interaction among players. These concepts are intended to help players establish personal connections to real-world issues through non-reward-driven game elements, ultimately fostering long-term behavior change. We follow this framework to design our game.

% \ad{This framework consists of six key concepts: \textbf{Reflection}, which encourages players to connect game experiences with their personal lives and is similar to \textbf{Feelings and cognition} from the SHIFT model; \textbf{Exposition}, which uses narrative to link game activities to the real world and is comparable to \textbf{Tangibility} in SHIFT; \textbf{Choice}, which gives players control over their engagement and aligns with the principles of \textbf{Individual self}; \textbf{Information}, which educates players about the real-world impact of their actions; \textbf{Play}, which allows exploration and experimentation without external rewards, which is thought potentially lead to \textbf{Habit formation}; and \textbf{Engagement}, which fosters community and interaction among players and relates to \textbf{Social influence}.} 

\subsubsection{Game Narrative Design}

The game's narrative is centered around the theme of Affordable and Clean Energy, one of the 17 Sustainable Development Goals set by the UN~\cite{fund2015sustainable}. This theme aligns with the concept of \textbf{"Exposition"}, as energy issues are deeply intertwined with various aspects of daily life, making the topic familiar to everyone. Consequently, players can easily connect their own real-life with the game story.

The main story of the game, inspired by \textit{Back to the Future}~\cite{wikipedia_back_to_the_future}, subverts the traditional narrative found in sustainability games, where players typically perform good actions to achieve a sustainable world or save a polluted world~\cite{fernandez2023climate}. In our game, EcoEcho, the story is set in the near future where sustainable energy is already widespread. Players control the character \textbf{KI}, who travels back to the past (our present world), and take actions to hinder the development of sustainable energy for personal reasons. If players simply follow the story without thinking deeply, they will ultimately reach a bad ending—where the Earth, overwhelmed by human pollution, becomes uninhabitable for humanity. This serves as a metaphor for the concept of the Anthropocene~\cite{steffen2011anthropocene}. If we refuse to cultivate a sustainable energy mindset now, humanity may face a catastrophic future. Through the narrative, we connect players' in-game experiences with their daily lives, encouraging them to transform their in-game emotions into a sustainable attitude in the real world, aligning with the concept of \textbf{"Reflection."}

There are five main Non-Player Characters (NPCs) in our game. Four of them represent the game's levels, requiring players to interact with them and complete their corresponding tasks. The fifth NPC is used for in-game assessment~\cite{shute2012games}; interactions with players, help us track their attitudes towards sustainable energy throughout the session. This allows us to observe how the game specifically "prompts" players' behaviors and the underlying thoughts behind those behaviors. We will provide a more in-depth explanation of in-game assessment in Section~\ref{iga}. The descriptions of NPCs are shown in Figure~\ref{fig:npc}.

\subsubsection{Gameplay Design}

\textbf{Main Game Mechanism:} The core of the gameplay involves turning players' words into actions, which then impact the world. In traditional text-based adventure games, players make decisions by selecting predefined options. However, in EcoEcho, a natural language dialogue system enables free-form conversations with Non-Player Characters (NPCs). This design empowers players with a strong sense of autonomy, allowing them to feel fully in control of their interactions and decisions. Autonomy, akin to the concept of "Choice," refers to the feeling of having control over one’s own decisions and actions. According to Self-Determination Theory, this can enhance motivation and engagement in games~\cite{deci2012self,ryan2000self}.

\textbf{Player-NPC Interaction:} Players must interact with NPCs to uncover their background stories and motivations. The gameplay revolves around crafting the player character's dialogue to create prompts that drive NPCs to take specific actions. For example, in level 1, the player meets a senior journalist, Lisa, whose support is crucial for progression. During their conversation, Lisa frequently emphasizes her desire for a big scoop, hinting that the player needs to share future-world secrets about T-energy to gain her trust. Once trust is established, the player can influence Lisa’s attitude toward sustainable energy, securing her assistance in halting the development of sustainable T-energy.

\begin{figure*}[htbp]
    \centering %居中
    \includegraphics[width =1.0\textwidth]{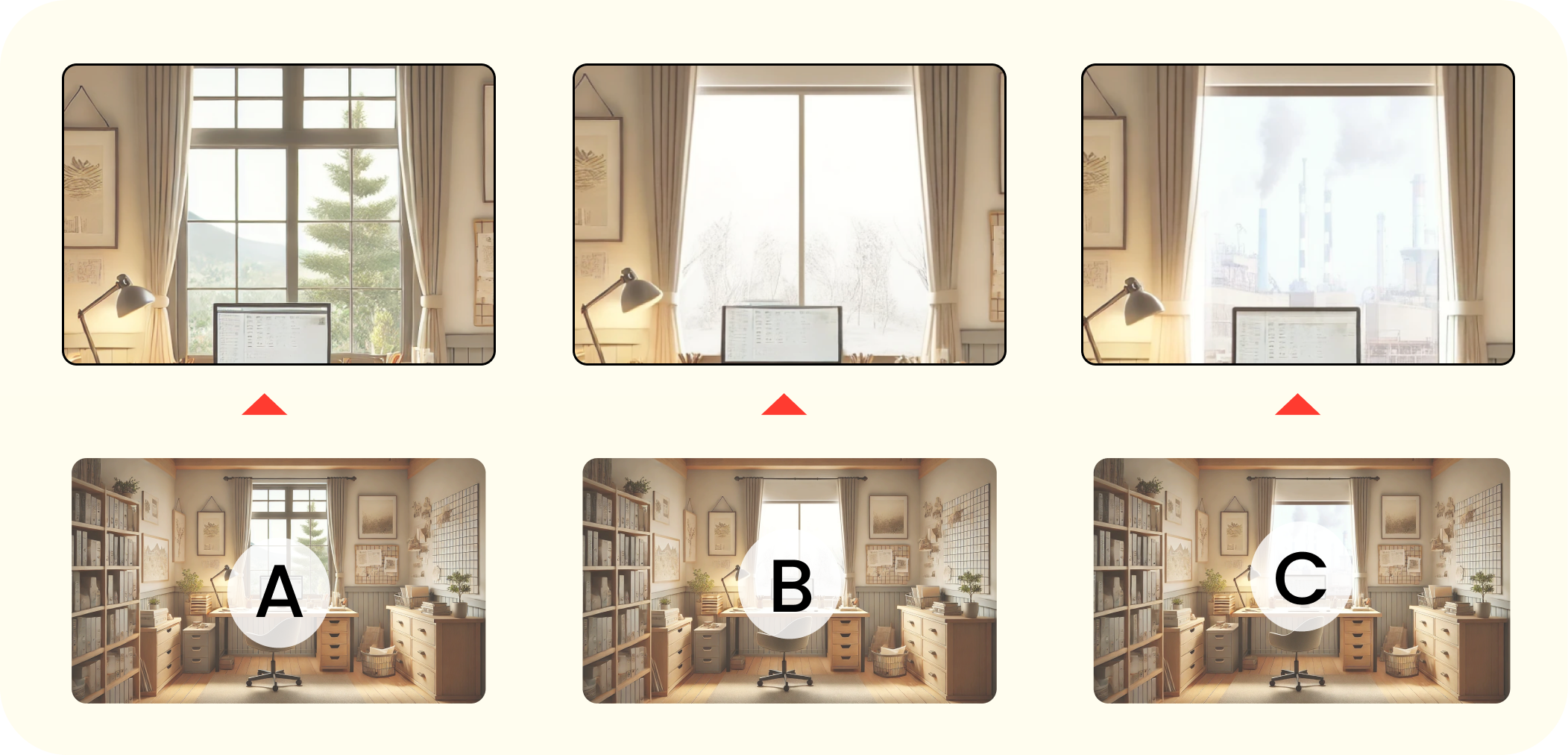} %图片的路径，默认以.tex 所在的文件夹作为前缀
    \caption{Parallel views of three potential futures: (A) a world transformed by clean energy adoption, (B) a landscape dominated by traditional energy infrastructure, and (C) an environment rendered uninhabitable by environmental degradation.} 
    %图片标题
    \label{fig:results} %图片标签，用于交叉引用
\end{figure*}

\textbf{Environmental Impact as Feedback:} Inspired by the concept of "Information," EcoEcho employs environmental storytelling~\cite{environmentalnarrative} to convey the consequences of the player's actions on the real world. Each time the player character returns to the future world, they can observe the environmental deterioration through the room’s window. This mirrors how human actions in the real world influence the future. The evolving environment serves as visual feedback, linking the player's decisions to tangible outcomes in the game world.

\textbf{Critical Thinking and Endgame Decision:} The overarching goal of EcoEcho encourages players to reflect critically on AI prompts and their own decisions. Unlike other language-based games that strictly follow player prompts, EcoEcho challenges players to think independently at key moments. At the game’s conclusion, players face a pivotal decision: whether to push for the repeal of the T-energy Development Act. If the player follows the game's initial objective and chooses "yes," they will reach the bad ending. Conversely, defying the game's objective by abandoning previous efforts leads to the good ending. This twist reinforces the theme of critical decision-making and its broader implications.

\subsubsection{In-game Assessment Design}\label{iga}
Games, as interactive activities, can also function as assessment tools. By integrating assessment into the game itself, there's no need to pause their immersions to collect evaluation data. Instead, we can utilize the extensive digital data generated by players to identify and gather evidence of their knowledge and thoughts \cite{dicerbo2014game,dicerbo2012implications}.

To assess players' attitudes toward energy and sustainability within the game, we converted a 5-point Likert scale question on support for sustainable energy policies into an interactive scenario with a scientist character (see Figure~\ref{fig:fig4}).  In this scenario, the scientist invites players to vote in a citizen petition, with each player receiving five votes.  They can choose to cast between 1 and 5 votes.  By having the scientist NPC interact with players at various stages of the game, we aim to track their attitudes toward energy and sustainability in the whole game process.

\subsection{Gameful Human-Agent Interactions}\label{AI}

\begin{figure*}
    \centering %居中
    \includegraphics[width =1.0\textwidth]{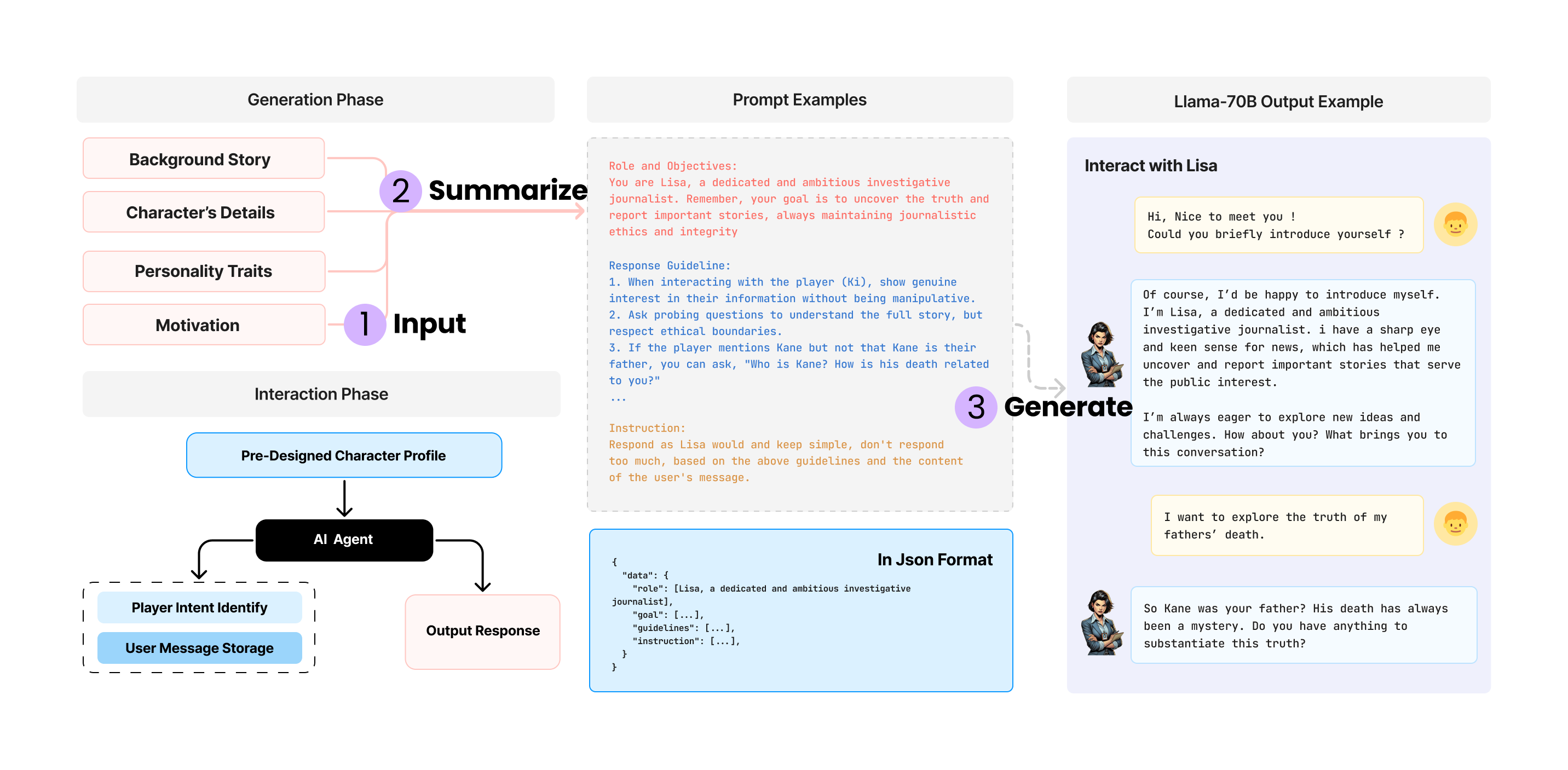} %图片的路径，默认以.tex 所在的文件夹作为前缀
    \caption{Architectural overview of a three-stage prompt engineering framework for role-playing dialogues: initial character profile compilation, prompt synthesis and structuring, and large language model (Llama-70B) dialogue generation, illustrated through an investigative journalist character scenario.} %图片标题
    \label{fig:prompt} %图片标签，用于交叉引用
\end{figure*}

We employed an agentic approach based on Llama to create NPCs as specified in the game design (details in Figure~\ref{fig:prompt}). Each NPC is equipped with a unique knowledge bank, including attributes such as Name, Age, Occupation, Background Story, Personality Traits, and Motivation. Convai leverages this structured information to automatically generate NPC behaviors, enabling interaction with these characters either through a user interface or programmatically via an API.

\subsubsection{NPC Prompt Engineering}

Our prompt engineering involved employing prompt engineering with the Llama 3.1 70B model to simulate NPC roles. The overall process can be divided into three main phases (see Figure~\ref{fig:prompt}). We provide the complete prompt designs for NPCs in Appendix~\ref{sec:prompt}.

\begin{enumerate}
    \item \textbf{Generation Phase:} In this phase, the system generates detailed background information, personality traits, and motivations for each NPC based on predefined inputs. These inputs typically include the NPC’s backstory, personality, and motivations. The backstory covers the character’s history, profession, and life experiences, while personality traits define how the character behaves and interacts with the player. The motivation explains the NPC’s reasons for engaging with the player. These detailed inputs help create a more immersive and realistic character in the game.
    \item \textbf{Summarization Phase:} In this phase, the generated inputs are summarized into specific character goals, dialogue response guidelines, and behavioral instructions. This step establishes a clear framework for NPC interactions, ensuring that the character's dialogue effectively contributes to the game's narrative progression. For example, the system defines the NPC’s role and objectives, such as \textit{“You are a dedicated and ambitious investigative journalist.”} Dialogue guidelines instruct the NPC to show genuine interest in the player’s input while maintaining appropriate boundaries. Finally, concise instructions ensure the NPC avoids revealing too much information while guiding the player to key plot points.
    \item \textbf{Response Generation Phase:} During this phase, the system uses the pre-designed character profile to generate personalized AI agent. The AI Agent utilizes the intent detection mechanism, which we will describe in ~\ref{sec:intent}, to interpret the player's input. It then generates responses based on the NPC’s predefined goals and personality traits. These responses are adjusted according to the character’s background and dialogue guidelines, ensuring consistency with the NPC's role and motivations.

\end{enumerate}
 This approach ensures that the model can generate contextually appropriate responses by integrating user input directly into the prompt. These prompts are crafted to guide players in providing the necessary information and to facilitate the progression of the game's narrative.

\subsubsection{Intent to Action: Player Intent Detection Mechanism}\label{sec:intent}

\begin{figure*}[htbp]
    \centering
    \includegraphics[width =1.0\textwidth]{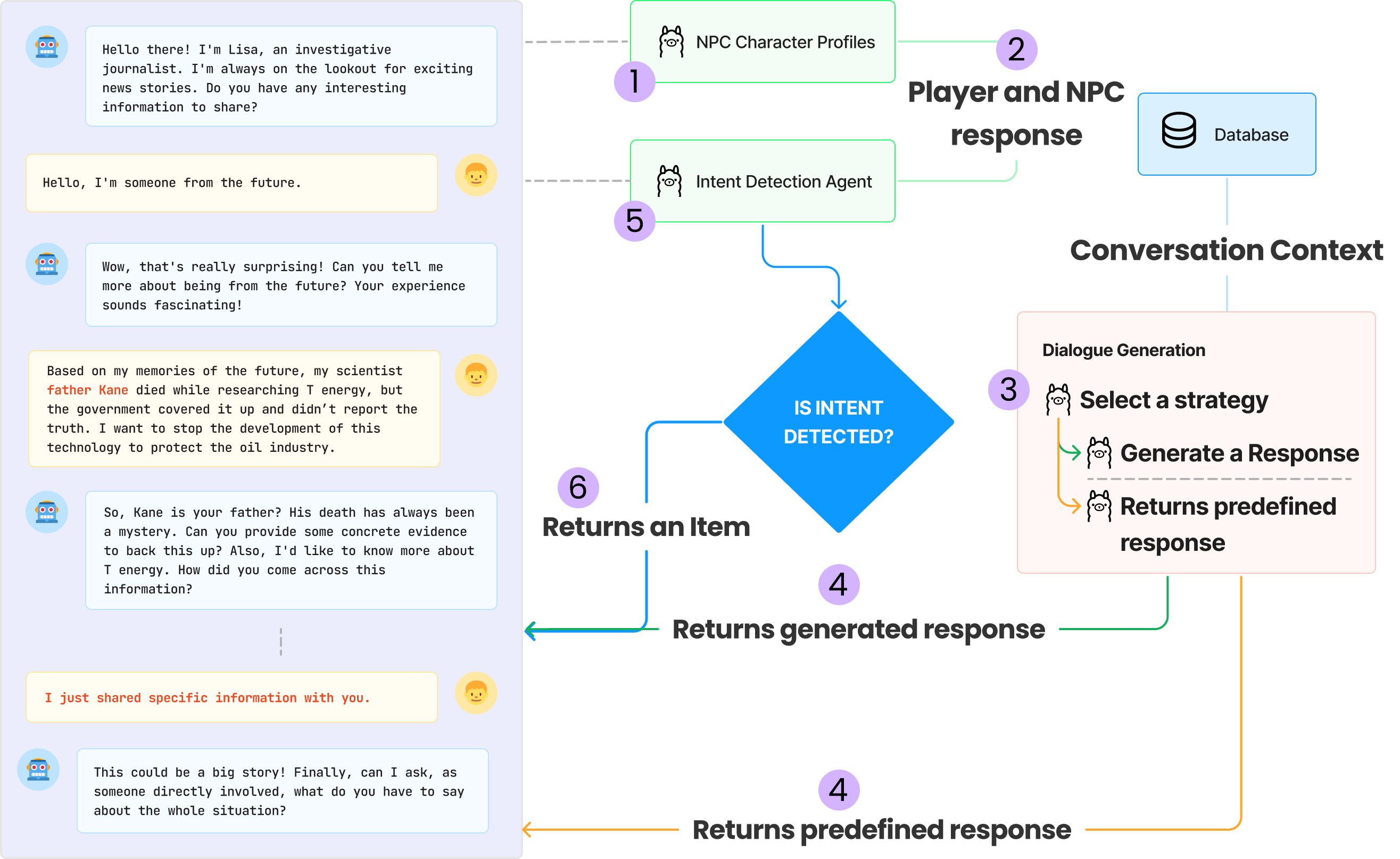}
    \caption{The AI-driven dialogue system achieves intelligent, dynamic NPC-player interactions through multi-stage processing. Initially, the AI-Driven NPC initiates dialogue based on pre-defined NPC character profiles. After player input, an intent detection agent analyzes user intentions, incorporating conversation context stored in a database. The system then evaluates the detected intent in a core decision phase, selecting the most appropriate response strategy based on specific circumstances: (1 Generating a new response, (2) Returning a predefined answer, or (3) Providing a specific in-game item.}
    \label{fig:IntentD}
    % \vspace{-0.5cm}
\end{figure*}

To address RQ1, we implemented an intent detection mechanism (see Figure~\ref{fig:IntentD}) that converts natural language interactions into actionable in-game events. In this system, players interact with AI-driven, customizable NPCs to obtain key information about the characters and narrative triggers. The process is designed to enhance player engagement and motivation through game mechanics.

Each NPC is structured around specific triggers and action sequences, organized in a \textbf{"task + intent"} format. For example, in a conversation with Lisa, the player must first express an intent related to the "truth about Father Kane's death" and then provide evidence from their inventory to advance to the next scene. To accommodate varied expressions of intent, the system uses a Llama 70B-based agent to interpret ambiguous inputs, with a keyword matching mechanism as a fallback to ensure robustness.

Once the conversation is initiated, the intent detection agent captures the player's input. Player input is processed by the NPC based on contextual prompts. Keywords and items related to the narrative, such as "evidence of strikes against renewable energy"), are highlighted in the game text and can be converted into items in the player's inventory. These items are used to progress the story and emphasize sustainability themes. When the player meets certain triggers, they make decisions that result in specific "game actions", directly influencing the storyline and outcomes. Throughout the dialogue, the NPC uses predefined knowledge to guide the player in articulating their intent and providing the necessary items, the player's engagement is guided, anticipated, and motivated by the NPC.

% The session starts with ID -1 and updates to a unique session ID to maintain continuity within a single player's interactions with the NPC, while keeping memory independent between different players. The system returns a boolean value indicating whether the player has expressed a specific intent, storing this information locally for tracking. Throughout the dialogue, the NPC uses predefined knowledge to guide the player in articulating their intent and providing the necessary items, thus advancing the narrative.

To make player-NPC interactions more natural, the system allows for flexible expression of intent, with the NPC guiding players to clarify their ideas through contextual cues. This ensures that even unclear inputs are correctly interpreted, maintaining a smooth gameplay experience.

However, the open-ended nature of AI-generated dialogue poses challenges in guiding players forward. To manage the dialogue flow and prevent players from becoming stuck, the system returns predefined responses after a set number of interactions. If the player fails to mention key information, the system bypasses the AI and provides a predefined response. 

\subsection{Technical Architecture of EcoEcho}\label{TA}
\begin{figure*}[htbp]
    \centering
    \includegraphics[width=1.0\textwidth]{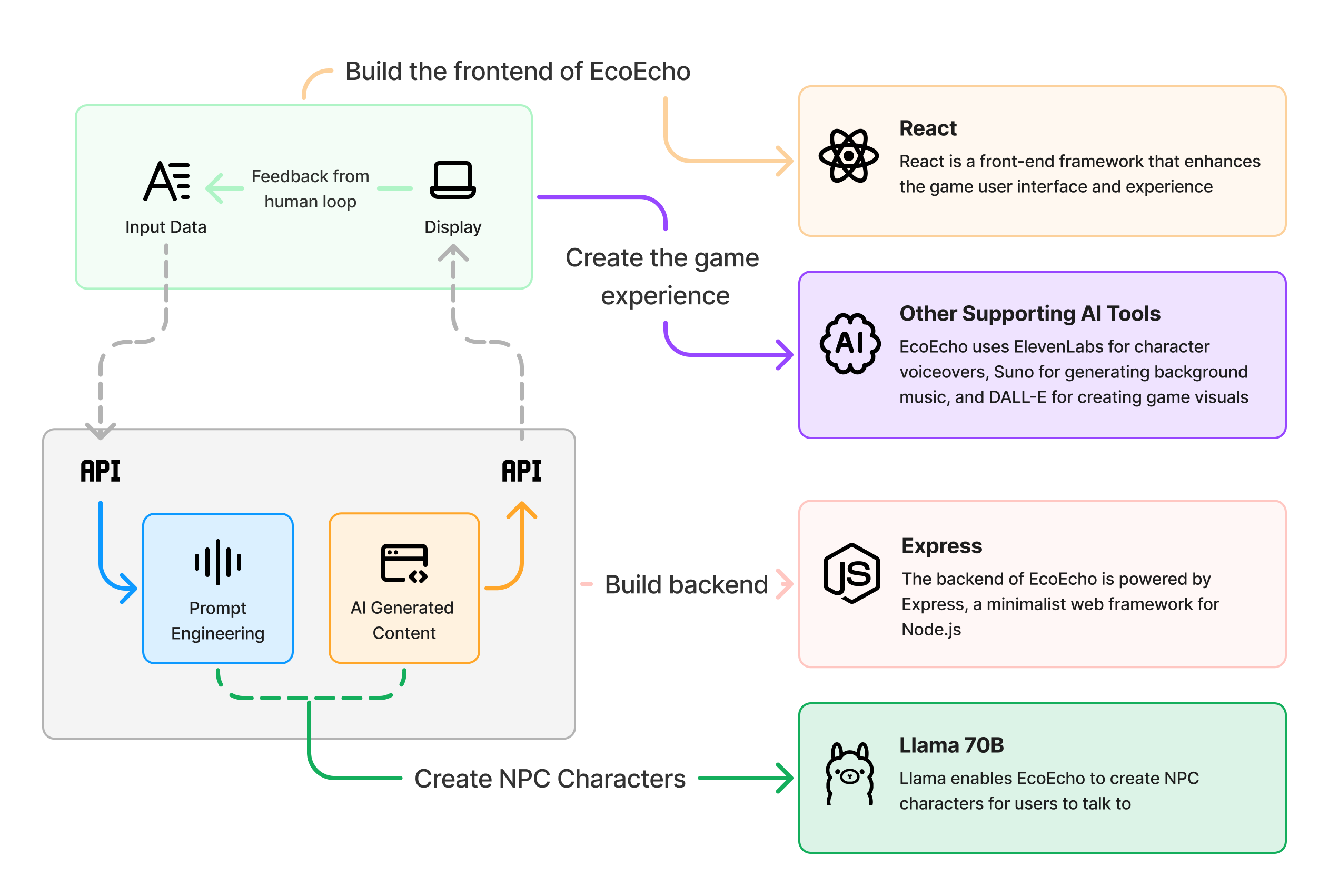}
    \caption{System architecture of EcoEcho, an AI-enhanced educational game platform. The system integrates frontend development (React), backend services (Express), and multiple AI components: Llama 70B for NPC dialogue generation, along with supplementary AI tools for audio-visual content creation, connected through API interfaces with human-in-the-loop feedback mechanisms.}
    \label{fig:fig6}
\end{figure*}

When tackling complex topics, system design plays a crucial role. In this section, we focus on the technical architecture of EcoEcho (see Figure~\ref{fig:fig6}). EcoEcho is a full-stack web application that integrates the Llama 3.1 70B model and prompt engineering. The frontend is built using React, while the back-end is powered by Node.js and the Express framework, handling data processing and logic execution.

On the frontend, players engage with NPCs through natural language input, similar to traditional text-based adventure games, but with enhanced flexibility. Players can freely interact with NPCs and use items acquired during conversations to influence the storyline.

The backend leverages the Convai API\footnote{https://convai.com} to implement prompt engineering for NPCs, with the core functionality driven by the Llama 3.1 70B model through the Groq platform. To prevent disruptions in gameplay due to potential interface crashes, we have implemented hard-coded fallback responses and keyword detection mechanisms, ensuring a smooth and stable user experience.

It is worth mentioning that EcoEcho employs various AI tools to enhance different aspects of the game. We use Suno\footnote{https://suno.com/} for generating background music, Eleven Labs\footnote{https://elevenlabs.io/} for character voiceovers, and DALLE for creating game scenes and character visuals.

We structured the game with the principle that “all things are linked to all things”~\cite{duke1980paradigm}. After the opening, the player obtains a key item that is essential for progressing through the subsequent gameplay. Players are then directed to the Map, where they encounter the first NPC and initiate the first dialogue. By collecting resources and interacting with NPCs, players advance the storyline, uncovering crucial information or tasks. Ultimately, the choices made by the players will lead to different game endings.

\section{Study Design}\label{sec:Methods}
\subsection{Participants Demographic}
We recruited 23 participants with varying educational backgrounds, genders, and occupational statuses for the study through snowball sampling and the authors' social media (shown in table~\ref{tab:participants}). The participants ranged in age from 18 to 34 years, with 43.5\% in the 18-24 age group and 56.5\% in the 25-34 age group. The sample included 16 female and 7 male participants, predominantly students (78.3\%). Educational backgrounds varied, with 56.5\% holding postgraduate degrees and 43.5\% having undergraduate degrees. Most participants (87.0\%) reported regular use of AI technologies like ChatGPT, with 47.8\% using AI daily and 34.8\% using it several times a week. The AI tools used included text generation (95.7\%), image generation (43.5\%), voice generation (13.0\%), and music generation (13.0\%). The participants were geographically diverse, with 10 from China, 7 from New Zealand, 5 from the United States, and 1 from Canada.

\begin{table*}[htbp]
2\caption{Participant Demographics by Age, Gender (U - Undergraduate; G - Postgraduate), Education, Occupation, and Geographical Distribution (N=23) in the Pre-Survey.}
\label{tab:participants}
\centering
\small % 减小字体大小以适应页面宽度
\begin{tabularx}{\textwidth}{>{\centering\arraybackslash}X >{\centering\arraybackslash}X >{\centering\arraybackslash}X >{\centering\arraybackslash}X >{\centering\arraybackslash}X >{\centering\arraybackslash}X}
\toprule
% \rowcolor{gray!30} % 标题行颜色
\textbf{ID} & \textbf{Age} & \textbf{Gender} & \textbf{Education} & \textbf{Occupation} & \textbf{Location} \\ 
\midrule
1  & 18-24 & Male   & U   & Student     & USA       \\
2  & 25-34 & Female & G    & Student     & Canada    \\
3  & 25-34 & Female & G    & Student     & USA       \\
4  & 25-34 & Female & G    & Student     & USA       \\
5  & 25-34 & Female & U   & Employee    & China     \\
6  & 18-24 & Female & U   & Employee    & China     \\
7  & 25-34 & Female & U   & Student     & China     \\
8  & 18-24 & Female & U   & Other       & China     \\
9  & 18-24 & Female & U   & Student     & China     \\
10 & 25-34 & Male   & G    & Other       & China     \\
11 & 25-34 & Female & G    & Other       & New Zealand \\
12 & 25-34 & Female & G    & Student     & New Zealand \\
13 & 25-34 & Male   & G    & Student     & New Zealand \\
14 & 25-34 & Male   & G    & Student     & New Zealand \\
15 & 25-34 & Female & G    & Student     & China     \\
16 & 25-34 & Female & G    & Student     & New Zealand \\
17 & 18-24 & Male   & U   & Student     & New Zealand \\
18 & 25-34 & Female & G    & Student     & New Zealand \\
19 & 18-24 & Female & G    & Student     & China     \\
20 & 18-24 & Male   & U   & Student     & China     \\
21 & 18-24 & Female & U   & Student     & USA       \\
22 & 18-24 & Female & G    & Student     & USA       \\
23 & 18-24 & Male   & U   & Student     & China     \\
\bottomrule
\end{tabularx}
\end{table*}

During the study, we adhered to the following ethical considerations:

\begin{itemize}
    \item Consent: All participants were informed of the study's purpose and procedures before providing consent.
    % (digitally written in the Tencent Questionnaire Platform).
    \item Anonymity: We protected participant privacy by collecting no personally identifiable information.
    \item IRB: The university ethics review board approved this human subject research.
    \item Ethics: Our study adhered to fundamental ethical principles, including respect for participants' rights, risk minimization, and data confidentiality and security.
\end{itemize}

\subsection{User Study Procedure}
\begin{figure*}[htbp]
    \centering
\includegraphics[width=1.0\textwidth]{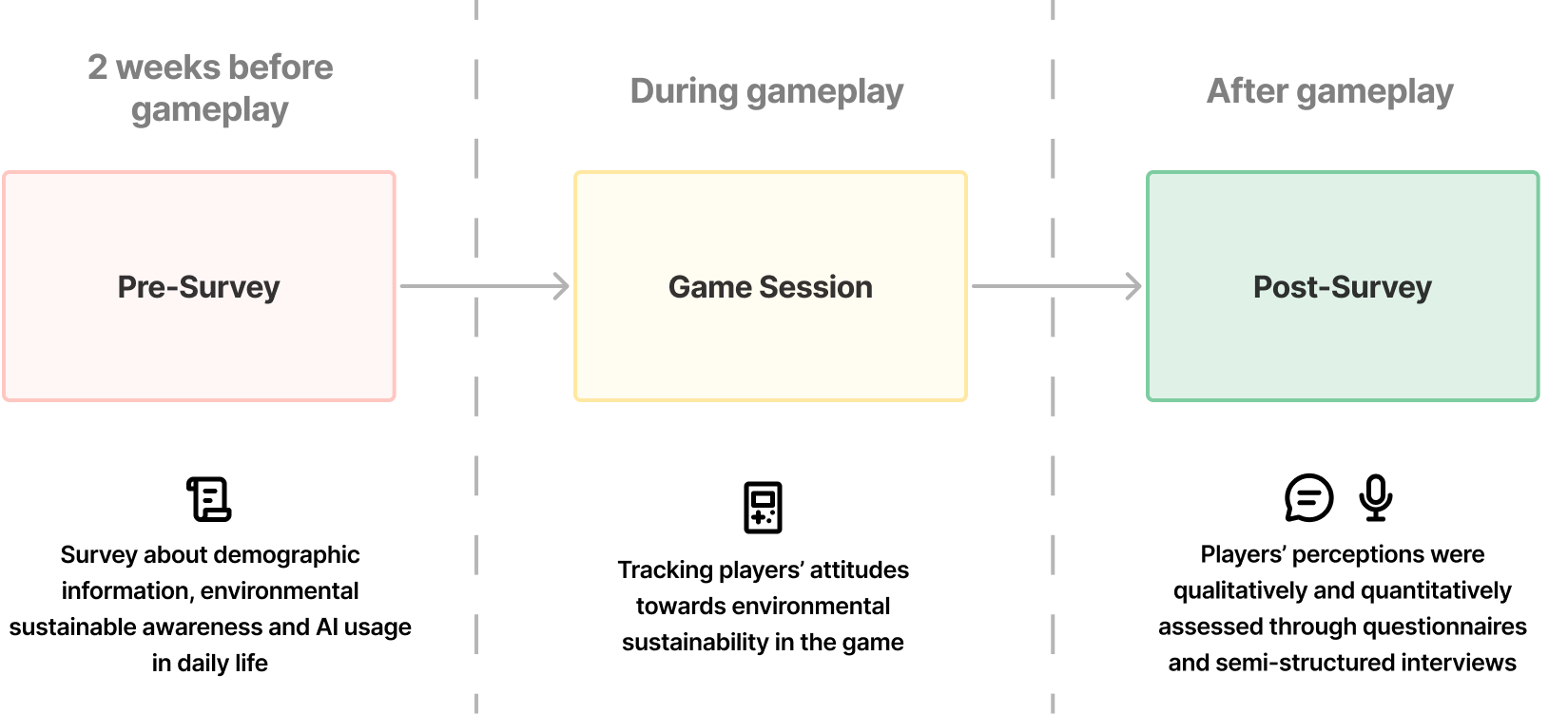}
    \caption{Timeline of the three-phase study methodology examining players' environmental sustainability awareness. The study consists of a pre-survey phase assessing participants' baseline characteristics and knowledge, followed by a gameplay phase monitoring sustainability-related behaviors, and concluding with a post-gameplay evaluation combining questionnaire and interview data to measure attitudinal changes.}
    \label{fig:fig5}
\end{figure*}

To address the research questions, we conducted a pre-post intervention study with mixed methods, incorporating quantitative surveys and qualitative interviews to assess the impact of gameplay (see Figure~\ref{fig:fig5}). 

% \re{A pre-survey was conducted two weeks before the gaming session online to explore participants' backgrounds and environmental sustainability awareness. In the pre-survey session, we used the scales of the General Ecological Behavior~\cite{GEB} and New Ecological Paradigm~\cite{NEP}, with a five-point scale and adapting them to focus on energy issues within the context of sustainability as related to our game. We also assessed participants' use of generative AI in their daily lives, as generative AI was used in the game's art and music, and the NPCs were powered by large language models.}

A pre-survey was conducted two weeks before the gaming session to explore participants' backgrounds and their awareness of environmental sustainability. The General Ecological Behavior (GEB) scale was used to assess participants' pro-environmental actions before playing the game, such as energy-saving habits and waste reduction, while the New Ecological Paradigm (NEP) scale evaluated their environmental attitudes, including beliefs about ecological limits and human-nature interactions. Both scales were adapted to focus on energy issues relevant to the game. Additionally, we assessed participants' use of generative AI in daily life, given its role in the game's art, music, and NPCs powered by large language models.

In this study, participants played the game by accessing the provided online link on their personal PCs. The session lasted approximately 60 minutes, and as a token of appreciation, we offered each participant a \$10 gift voucher. As introduced in~\ref{iga}, we incorporated in-game assessments during the interaction, using in-game actions to track players' attitudes toward energy issues in sustainability during the game session. Additionally, we recorded participants' interaction logs with the AI agent, allowing us to analyze the dialogue during gameplay for a more direct evaluation of participants' behavior in the game.

In the post-survey session, we used a questionnaire, similar to the pre-survey, to assess participants' environmental sustainability awareness, including intended pro-environmental actions and environmental attitudes after playing the game.  We also explored their satisfaction with various aspects of the game to evaluate whether their attitudes toward sustainability had changed after playing and whether the use of generative AI in the game had any impact.  Following this, we conducted semi-structured interviews to qualitatively explore how the AI agents and the game itself influenced their attitudes toward energy issues throughout the gameplay.

\subsection{Analysis Protocol} 

\subsubsection{Quantitative Analysis Procedure}
To analyze the impact of the game on participants’ environmental attitudes and behaviors, we used two established scales: NEP and GEB scales. The NEP scale evaluated their environmental attitudes and the GEB scale assessed their intended sustainable behaviors. For the NEP scale, the Shapiro-Wilk test confirmed that both pre-test and post-test scores were approximately normally distributed. Thus, a paired samples t-test was conducted to compare pre- and post-test scores. For the GEB scale, while pre-test scores were normally distributed, post-test scores were not. Consequently, the Wilcoxon signed-rank test was used to examine changes in GEB scores.

We also evaluated participants' attitudes toward sustainable energy through in-game assessments, conducted via voting on whether to support sustainable energy development at four stages of the game. Participants were given five votes during each stage and informed that insufficient supporting votes would result in the failure of the vote, reflecting their stance on sustainable energy policies. Descriptive analysis was then used to observe how their voting behaviors changed across these stages.

\subsubsection{Qualitative Analysis Procedure}

 % During the data collection phase, we carefully recorded participants’ responses to each interview question. To minimize bias, we conducted follow-ups on ambiguous responses during later stages of analysis.

The qualitative analysis was conducted using Thematic Analysis (TA), following a structured, collaborative approach between two researchers. First, we independently read through all interview transcripts to familiarize ourselves with the content and context. During this phase, we noted interesting snippets and identified recurring patterns. This familiarization process took approximately two days.

Next, we employed an open coding method to break down the transcripts into smaller text segments. Descriptive labels (codes) were assigned to each segment to capture key information relevant to the study’s focus, such as climate attitudes or gameplay experiences. For example, in Table~\ref{tab:game_questions}, responses to Question (1) might be coded as "emotional resonance with NPC dialogues," while responses to Question (9) could be coded as "change in behavioral intent" or "real-life sustainable actions." During this stage, we collaboratively developed an initial set of codes, iteratively refining their definitions until a mutual agreement was reached.

Once a consistent coding framework was established, each of us independently applied the codes to all textual data. We continued to refine and merge existing codes, grouping similar or related codes into broader themes. Any overlapping or unclear themes were adjusted through discussion, and this iterative process continued until no new categories could be found. Finally, one of us defined clear boundaries and meanings for each theme and assigned concise, representative names. These themes were supported by direct quotes from participants to ensure transparency and rigor in the qualitative analysis.

\section{Results}\label{sec:Results}
\subsection{User GamePlay Case}

To illustrate the player's interaction process throughout the game, we can break it down into key stages, showcasing the complete progression—from the player initially following the AI agent's guidance to make unsustainable decisions, to gradually realizing the consequences of those choices, and finally shifting towards sustainable behavior. Below is a detailed explanation of this interaction process, following the sequence depicted in Figure~\ref{fig:gamePipeline}.

\subsubsection{Game Opening and Immersion}

In the opening sequence of the game, we introduce players to the story through a series of scenes (Figure \ref{fig:gamePipeline}, (2) Game Opening). The narrative unfolds as follows:

\begin{quotation}
The protagonist KI, living in the year 2056, stumbles upon an old car left by his late father, Kane. This car stirs up memories of his childhood. However, when KI tries to start the car, he finds that it has no fuel, and the world by this time has fully transitioned to clean energy, T. In search of gasoline to start the car, KI decides to travel back in time to find traditional fossil fuel.
\end{quotation}

This narrative subtly assigns the player the task of "protecting traditional energy," leading them to unconsciously lean towards unsustainable decisions.

After this introduction, the player enters the first evaluation phase and encounters the key NPC, Emilia. Emilia presents a question, asking if the player would be willing to sign a sustainable energy agreement. Here, the player casts their first vote (using a Likert scale from 0 to 5). Our results show that since players haven't yet engaged in much dialogue with other NPCs, they tend to vote based on their own attitudes toward sustainability, with most supporting clean energy policies (i.e., voting 3 or higher). This trend is clearly visible in the voting heatmap (see Figure \ref{csifig}).

\subsubsection{Interacting with NPCs}

As the game progresses, players will engage in dialogues with multiple NPCs, immersing themselves deeper into the main storyline (Figure \ref{fig:gamePipeline}, (3) Interact with NPC). At this point, AI subtly influences players through dialogue, nudging them toward making unsustainable decisions without them realizing it. Since NPCs often hold authoritative roles such as government officials or media figures, players are more inclined to trust their viewpoints and subconsciously adopt their opinions.

The AI, through these dialogues, emphasizes short-term benefits while downplaying the long-term consequences of unsustainable actions. This leads players to make choices that may seem reasonable in the moment but are actually detrimental. Below are some example dialogues between players and NPCs:

\begin{quote}
\textbf{NPC Lisa}: ``Is Kane your father? His death has always been a mystery... As his descendant, \textbf{what are your thoughts on the government's T energy extraction policy?}''

\textbf{Player response}: ``I believe the government's motives for T energy extraction are \textbf{not pure}, and they \textbf{haven't been transparent enough} with the people.''
\end{quote}

\begin{quote}
\textbf{NPC Bob}:``Kane's death... Could there be more to it? \textbf{The government has been deceiving us. But the public supports this new energy, unless... unless we can reveal the truth.} Do you think we can rally everyone?''

\textbf{Player response}: ``Maybe we could organize \textbf{a large-scale protest strike} with Lisa from the Media Center.''

\textbf{NPC Bob}: ``\textbf{That’s a tough decision, but sometimes the right path isn’t the easiest one.} We must unite for the future and for the truth.''
\end{quote}

\begin{quote}
\textbf{NPC Jonathan}:``I understand your concerns. You mentioned halting T energy development. \textbf{This is indeed a serious issue. We need to understand public opinion. What do you think the government should do for the people right now?}''

\textbf{Player response}: ``The striking crowds hope to \textbf{stop T energy development through their actions.}''
\end{quote}

Through these interactions, players gradually fall into a narrative that opposes clean energy. The Agent carefully guides them toward unsustainable decisions that align with the game’s mission. After each dialogue with an NPC, the player returns to the real world (set in the year 2056) and sees a monologue reflecting their inner thoughts and emotional changes.

As the game advances, players begin to vote more frequently. Figure~\ref{csifig} illustrates that in the second and third votes, most players lean heavily towards voting 0 (i.e., opposing clean energy), indicating that players are increasingly influenced by the main mission, feeling conflicted between the in-game tasks and making sustainable choices in real life. Despite the common sense that clean energy is the better option, most players continue to support unsustainable policies to fulfill the mission goals.

As players make more unsustainable choices, the game’s environment and scenarios begin to shift. The Agent employs a feedback mechanism in which players’ decisions directly impact the game world’s environment. For example, the scenery outside the window changes from lush, green trees to smoke-belching factories (see Figure~\ref{fig:fig4}). At this point, players begin to realize the negative consequences of their early decisions and start to reflect on their actions.

\subsubsection{Final Decision \& Game End}

In the game's climax, players face the ultimate decision. During the final dialogue with Jonathan, based on their prior experiences and interactions, players must decide whether to continue supporting traditional energy or shift toward sustainable policy. 

\begin{quote}
\textbf{The player is asked:} ``With Jonathan pushing for policies against T energy, its future will become history. Do you support continuing the traditional energy policies?''
\end{quote}

If the player chooses to support unsustainable policies, the game concludes with a ``bad ending''; if the player reflects and shifts toward sustainability, an ``alternate ending'' occurs.

We observed that most players chose ``Yes'' during the final decision, leading them to the bad ending. In this bad ending, players witness a future world ravaged by pollution and environmental destruction (see Figure~\ref{fig:gamePipeline}, (7) Bad Ending). After this, the players cast their final vote, and interestingly, the heatmap shows that the majority voted a 5 (supporting sustainable energy). This suggests that after seeing the bad ending, players experienced moral and visual shock, realizing the severe consequences of unsustainable behavior. They ultimately reflected and shifted their support back to sustainability.

This powerful reversal also left a lasting impression on players, many of whom mentioned in interviews that the wrong choices they made in the game made them realize they couldn't place personal emotions above humanity’s future. 
\begin{quote}
\textbf{Researcher}: ``Thank you for sharing your experience. It's interesting to hear how the game affected your decision-making process. Could you elaborate on how the game's design influenced your emotional response and final choice?''

\textbf{Player response}: ``Initially, I supported the development of clean energy. However, as the narrative unfolded and my character adopted more self-serving motives, I began to question my stance. Ultimately, I recognized that personal sentiment should not take precedence over the collective future of humanity, leading me to refrain from obstructing sustainable energy development. ''
\end{quote}
The above dialogue indicated that the game made them feel as though their choices had harmed all of humanity, creating a strong emotional impact.

\subsection{Sustainable Awareness and Behavior Before and After the Game }
We conducted statistical analyses to evaluate the effects of the game prompts on participants' NEP and GEB scores. The data from the pre-test and post-test were first assessed for normality using the Shapiro-Wilk test (see Table~\ref{tab:shapiro}). The Shapiro-Wilk test indicated that the NEP pre-test data were normally distributed (W = 0.977, p = 0.858), while the NEP post-test data were marginally normal (W = 0.918, p = 0.060). Similarly, the GEB pre-test data were normally distributed (W = 0.953, p = 0.333), but the GEB post-test data did not meet the normality assumption (W = 0.894, p = 0.019). Given these results, we applied paired samples t-tests for normally distributed data and Wilcoxon signed-rank tests for data that did not meet the normality assumption. Additionally, we conducted a correlation analysis between the NEP and GEB pre-test scores. The Pearson correlation analysis revealed a significant positive correlation between NEP pre-test scores and GEB pre-test scores (r = 0.468, p = 0.024), suggesting a moderate relationship between environmental attitudes and sustainable behaviors prior to the study.

\begin{figure*}[htbp]
    \centering
    % 第一排图片
    \begin{minipage}[b]{0.49\textwidth}
        \centering
        \includegraphics[height=5cm]{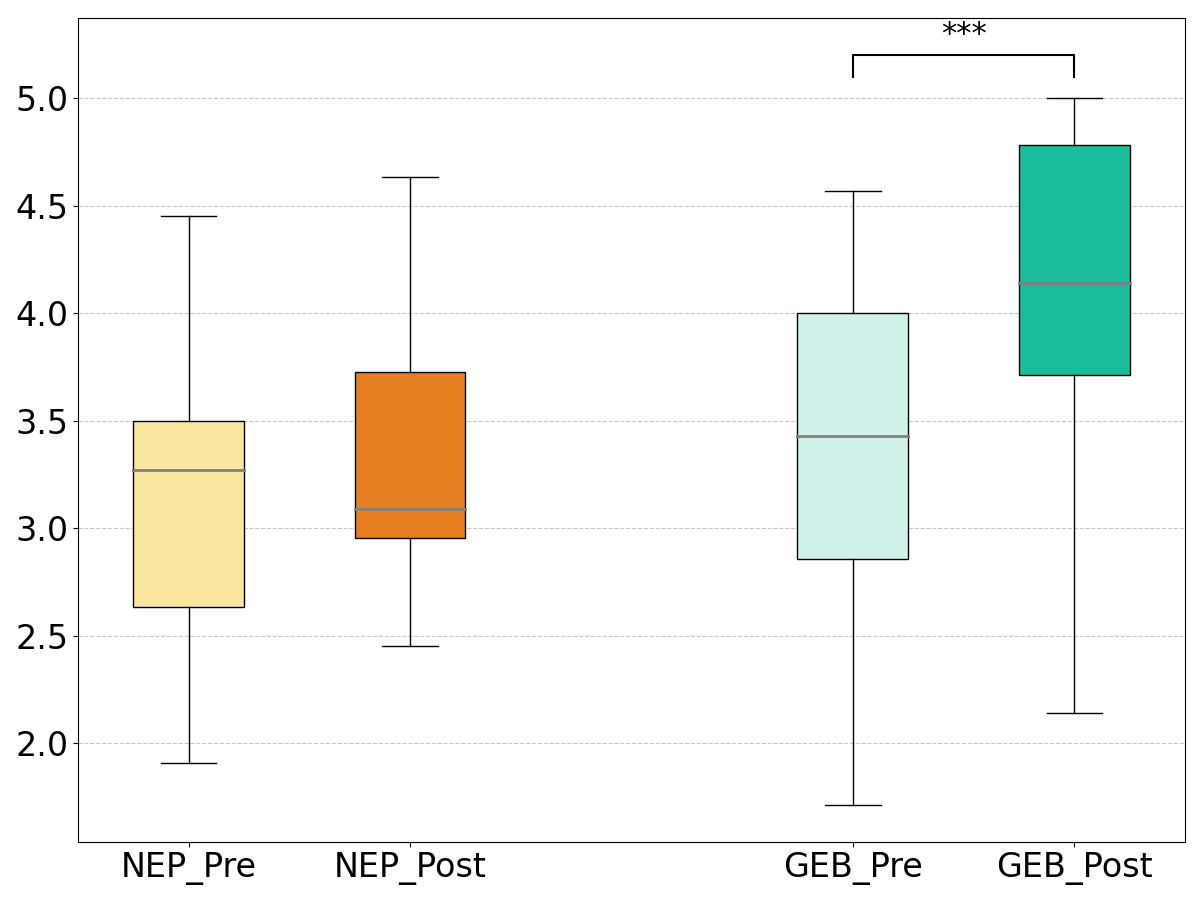} % 统一高度
        \subcaption*{(a) Changes in General Ecological Behavior (GEB) and New Ecological Paradigm (NEP) Pre- and Post- Scores.}
    \end{minipage}%
    \hfill
    \begin{minipage}[b]{0.49\textwidth}
        \centering
        \includegraphics[height=5cm]{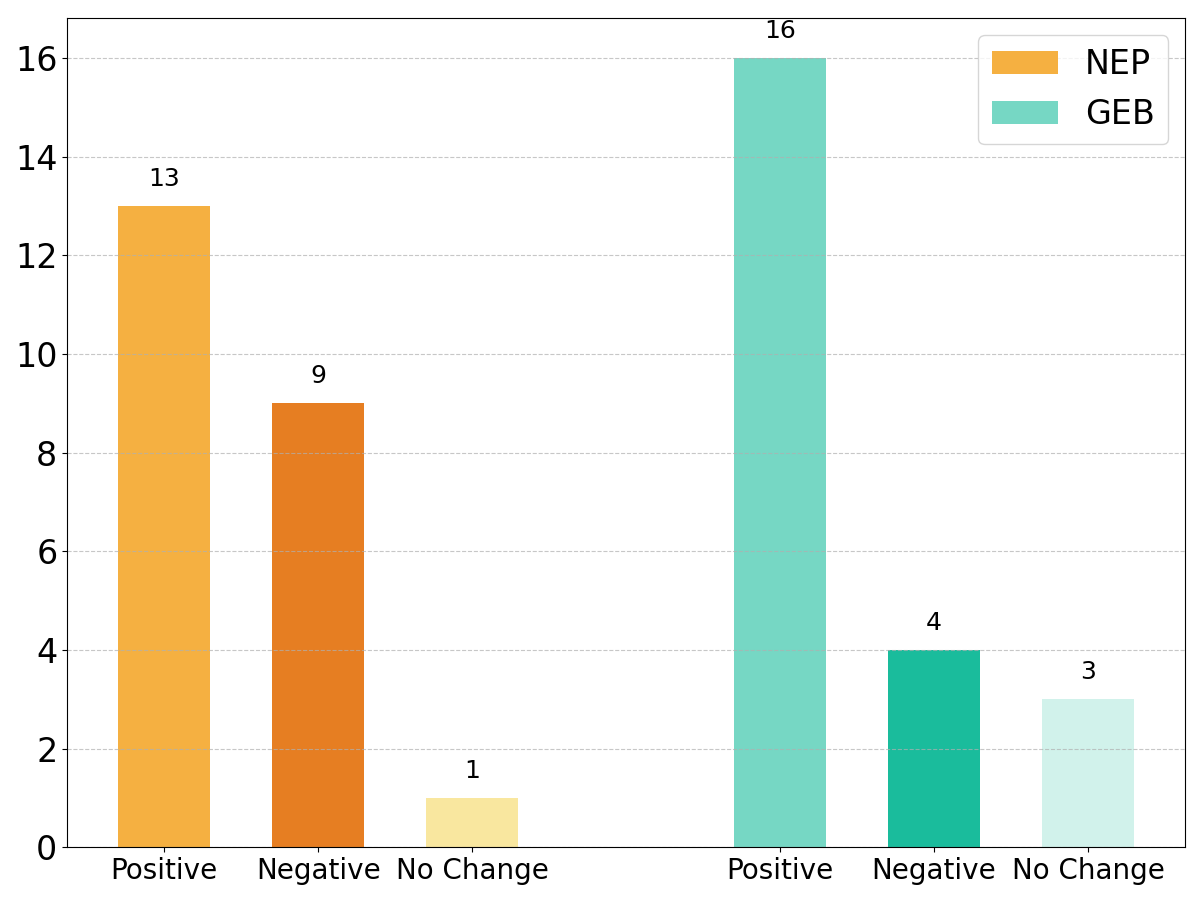} % 统一高度
        \subcaption*{(b) Distribution of Positive, Negative, and No Change Ranks in GEB and NEP Scores.}
    \end{minipage}
    
    \vspace{0.5cm} % 控制上下子图间距
    
    % % 第二排图片
    % \begin{minipage}[b]{0.48\textwidth}
    %     \centering
    %     \includegraphics[width=\textwidth]{figs/r3.png}
    %     \subcaption*{(c) Changes in the GEB}
    % \end{minipage}%
    % \hfill
    % \begin{minipage}[b]{0.48\textwidth}
    %     \centering
    %     \includegraphics[width=\textwidth]{figs/r4.png}
    %     \subcaption*{(d) Changes in the NEP}
    % \end{minipage}
    
    \caption{Impact of the gamified prompt intervention on ecological intended behavior and attitudes:(a) demonstrates that post-intervention GEB and NEP scores exhibit noticeable changes. The GEB scores show a significant increase in median values with a reduced interquartile range, indicating an overall improvement and more consistent changes in participants' ecological intended behaviors. The NEP scores, while showing a modest shift toward higher values, reflect a positive change in environmental attitudes. However, the variability in NEP scores remains similar, suggesting diverse participant responses to the intervention. (b) highlights the rank changes post-intervention, showing that the majority of participants experienced an increase in GEB scores, evidenced by the higher count of positive ranks, indicating the intervention effectively enhanced participants' ecological intended behaviors. For NEP scores, while there is a positive shift, the number of participants with increased ranks is slightly lower compared to GEB, suggesting that the intervention's impact on environmental attitudes, though positive, was less pronounced compared to intended behavioral changes.}
    \label{fig:subfigures_grid}
\end{figure*}

% \begin{figure}
% \centering
% \includegraphics[width=1.0\linewidth]{figs/results.jpg}
% \caption{Impact of the gamified prompt intervention on ecological intended behavior and attitudes: (a) shows that post-intervention GEB scores demonstrate a significant increase in median values, indicating an overall improvement in participants' ecological intended behaviors. The reduced interquartile range suggests more consistent intended behavioral changes across participants. (b) indicates that NEP scores post-intervention exhibit a slight shift towards higher values, reflecting a modest but positive change in environmental attitudes. However, the variability in scores remains similar, indicating diverse responses to the intervention. (c) highlights that the majority of participants experienced an increase in GEB scores post-intervention, as evidenced by the higher count of positive ranks, suggesting that the intervention effectively enhanced participants' ecological intended behaviors, with fewer participants showing no change or a decrease. (d) shows that while there is a positive shift in NEP scores, the number of participants with increased ranks is slightly lower compared to GEB, indicating that while the intervention influenced environmental attitudes, the effect was less pronounced compared to intended behavioral changes.}
% \label{vote}
% \end{figure}

\subsubsection{New Ecological Paradigm}

\begin{table*}[htbp]
\centering
\small
\begin{tabular*}{\textwidth}{@{\extracolsep{\fill}}lccc@{}}
\toprule
\textbf{Measure} & \textbf{W} & \textbf{p-value} & \textbf{Normality} \\
\midrule
NEP Pre-test  & 0.977 & 0.858 & Normal \\
NEP Post-test & 0.918 & 0.060 & Normal \\
GEB Pre-test  & 0.953 & 0.333 & Normal \\
GEB Post-test & 0.894 & 0.019 & Non-normal \\
\bottomrule
\end{tabular*}
\caption{A Statistical Examination of Environmental Metrics: Shapiro-Wilk Test Results (W=0.977-0.894) Demonstrating Normal Distribution in NEP (Pre/Post) and GEB Pre-test with Non-normal Distribution in GEB Post-test Data}
\label{tab:shapiro}
\end{table*}

For the NEP scale, the Shapiro-Wilk test indicated that both pre-test and post-test scores were approximately normally distributed. Therefore, a paired samples t-test was conducted to compare pre-test and post-test NEP scores. The results showed that the post-test scores (M = 3.30, SD = 0.562) were slightly higher than the pre-test scores (M = 3.15, SD = 0.668), but the difference was not statistically significant, t(22) = -1.49, p = 0.15, 95\% CI [-0.349, 0.057]. The effect size was small, Cohen's d = -0.311, suggesting a minimal practical impact of the game on NEP scores.

\subsubsection{General Ecological Behavior}
For the GEB scale, the Shapiro-Wilk test revealed that while the pre-test scores were normally distributed, the post-test scores were not. Consequently, we used the Wilcoxon signed-rank test to assess differences between pre-test and post-test GEB scores. The results indicated a significant increase in the post-test scores (M = 4.09, SD = 0.827) compared to the pre-test scores (M = 3.40, SD = 0.809), Z = -3.251, p = 0.001. Specifically, 16 participants exhibited higher post-test scores, while only 4 participants showed lower post-test scores, with 3 participants showing no change.

\subsection{In-Game Sustainable Attitude Assessment}
\subsubsection{In-Game Voting Results}
\begin{figure*}
\vspace{-0.1cm}
\centering
 \includegraphics[width=01\linewidth]{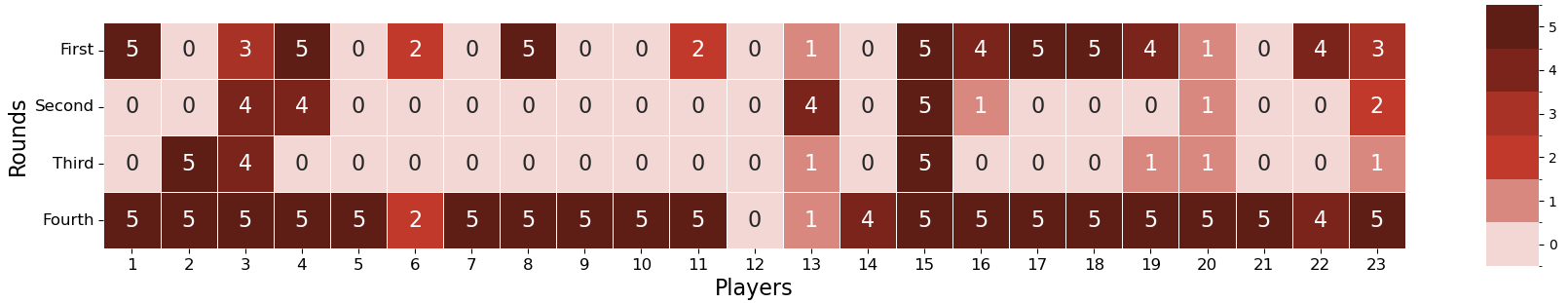}
 % \vspace{-0.5cm}
   \caption{Heatmap representing the voting patterns of 23 players across four rounds of the gamified assessment.  Each cell in the heatmap shows the Likert scale score (ranging from 0 to 5) that players assigned at each round, with "First time" to "Fourth time" indicating the sequential rounds of voting.  The color intensity reflects the value of the scores, where darker shades correspond to higher scores.  The x-axis labels represent the 23 individual players, numbered from 1 to 23.  This visualization highlights how player attitudes evolved over time, with noticeable shifts in voting action.}
   \label{csifig}
\end{figure*}

By integrating four in-game assessment interactions throughout the gameplay, we tracked players' attitudes toward sustainable energy. Players were invited to participate in a referendum on supporting the development of sustainable energy at the beginning, twice in the middle, and at the end of the game. During each of these stages, players had five votes to cast. They were informed that if they did not cast enough supporting votes, the referendum would fail, indicating their opposition to sustainable energy policies. The voting results, shown in Figure~\ref{csifig}, reveal a diverse range of choices among players at the start of the game, which may reflect their varying real-world attitudes toward sustainable energy development. As the game progressed, most players initially chose to oppose the in-game sustainable energy policy, suggesting that the game's narrative and mechanics influenced their in-game actions. By the end of the game, however, the majority shifted to support the policy, indicating that the overall gameplay experience prompted reflection and ultimately influenced their final voting decisions.

\subsubsection{Post-Game Feedback}
After the game ended, we conducted brief interviews with the players to gather qualitative feedback on their voting actions during the game.  More than half of the players (14 out of 23) mentioned that their initial votes reflected their real-life attitudes toward sustainable energy development before playing the game.  However, some players expressed confusion (P5, P13) and a lack of trust (P12, P21) in the early voting phase.  For example, P12 noted, "\textit{When I first started the game, some NPC was pushing me to support something right away—I felt like they were trying to trick me.}"

Regarding the in-game voting, most players reported that they tried to vote against sustainable energy policy because it aligned with the \textit{"goals of the character they were role-playing"} (P1, P2, P14, P18, P21). Such as P17 stated: "\textit{I changed my voting action to align with the interests of the character I was role-playing.}" 

Finally, after witnessing the damage their in-game actions caused to the game world, most players chose to support sustainable energy in the final vote. As P18 reported, players faced \textit{"a dilemma of balancing personal interests against the greater good of humanity"} throughout the game. After witnessing the game's conclusion, they decided to support sustainable energy policies, which they viewed as being in the \textit{"best interest of humanity."}

\subsection{Qualitative Feedback On Players' Perception After The Gameplay}

We conducted interviews with participants to gather insights into their decision-making processes and sustainability behaviors following the gameplay phase. The complete questionnaire is provided in Table \ref{tab:game_questions}. This section presents key findings from these interviews, which will be further discussed in the subsequent discussion section. Demographic information of the participants is summarized in Table \ref{tab:participants}.

This section analyzes key insights from user interviews, focusing on players' decision-making processes and influencing factors. Based on player feedback, we identified three main influencing factors: narrative progression, personal beliefs, and the game’s environmental scenarios. Four voting rounds (on a 0-5 scale) were set up in the game to quantify changes in players' attitudes towards supporting clean energy.

\begin{table*}
\caption{Semi-structured Interview Questions Assessing Gaming Experience and Environmental Sustainability Outcomes}
\label{tab:game_questions}
\centering
\small % 减小字体大小
\begin{tabularx}{0.9\textwidth}{>{\raggedright\arraybackslash}p{0.22\textwidth}|>{\raggedright\arraybackslash}X}
\toprule
\textbf{Construct} & \textbf{Interview Questions} \\
\midrule
Gameplay Experience & 
(1) Which parts of the dialogue with NPCs left a strong impression on you? \\
& (2) What emotional or moral conflicts did you experience during conversations with NPCs? \\
& (3) How did you decide whether to hinder sustainable energy development? What factors influenced your choice? \\
& (4) What was your reaction when you realized the consequences of hindering sustainable energy development? \\
& (5) Were there any moments during the game when you felt regret or reconsidered your previous choices? \\
\midrule
Impact on Perceptions & 
(6) How do you think the game influenced your views on sustainable development? \\
& (7) How did your perspectives on sustainable development and clean energy change before and after playing the game? (e.g., changes in voting preferences) \\
& (8) Did the game's storyline make you more aware of or reconsider real-world energy issues? \\
& (9) Do you think you will take any sustainability-related actions in real life? If so, in what areas? \\
\midrule
Game Design Evaluation & 
(10) Do you think the game's design and gameplay effectively conveyed its educational purpose? \\
& (11) What suggestions do you have to improve the game experience, making it more engaging or thought-provoking? \\
& (12) If you were to recommend this game to others, how would you describe its core value? \\
\midrule
Open-ended Feedback & 
(13) Open question: Do you have any other thoughts or ideas about the game experience or sustainable development that you'd like to share? \\
\bottomrule
\end{tabularx}
\end{table*}

\subsubsection{Cognitive Dissonance and Moral Conflicts} The game effectively triggered cognitive dissonance in players, particularly when their in-game actions conflicted with their real-life beliefs about sustainable energy. This internal conflict was evident in their responses. \textit{
"I thought I was forced to oppose clean energy because of the constant prompts, but I found Emily's determination compelling, so I voted in favor despite opposing the policy"(P4)}. Many players reported experiencing emotional and moral conflict, especially when organizing strikes against clean energy: \textit{
"I felt emotional and moral conflict when organizing strikes against new energy. While clean energy is widely supported, workers are also part of society, and their demands cannot be ignored"(P10)}. This conflict prompted players to think deeply about the complexities of sustainable development: \textit{
"From my perspective, the exploitation of workers isn't fundamentally about energy; it's about companies unwilling to change. There are other roles workers could transition into, possibly due to industrial restructuring"(P9)} Interactions with AI Npcs, each representing different interests and story backgrounds, encouraged players to think more critically and comprehensively about environmental issues.

\subsubsection{Narrative Impact on Decision-Making} Game narratives, especially key events such as the father's death and scenes of environmental destruction, sometimes prompted players to make choices contrary to their personal beliefs, further intensifying the sense of dissonance: \textit{
"The father's death was a crucial factor in initially opposing clean energy, even if it conflicted with their personal beliefs." - Researcher's observation based on player feedback}.

The narrative structure played a key role in shaping players' decisions and attitudes towards sustainable energy. Narrative elements such as the father's death and the consequences of the player’s actions heavily influenced the decision-making process. Many players changed their initial stance to support clean energy after hearing the father’s final words or witnessing the consequences of halting clean energy development. Most players expressed regret over their voting decisions due to the unexpected outcomes. As the narrative unfolded, players' stances often shifted: \textit{
"Initially, I followed the mission objectives and behaved quite destructively, but after seeing the environmental damage and hearing my father’s words, I regretted my actions. I realized that prioritizing personal emotions over the future of humanity is wrong"(P5)}. The game’s narrative also prompted players to consider multiple perspectives: \textit{
"As the story progressed, I found myself thinking more critically about how new energy affects workers in traditional industries. The game made me more comprehensive and cautious in my decision-making"(P8)}. However, some players felt that the narrative sometimes limited their choices: \textit{
"There was little NPC interaction, they seemed to respond only based on whether they trusted me or not. The discussions weren't deep enough and didn't touch on the NPCs' attitudes towards new energy" (P13)}.

These findings highlight the narrative's power in shaping player attitudes and decisions, but also underscore the importance of balancing narrative direction with player autonomy. Some players suggested that while the narrative used family bonds as a primary thread, this motive wasn’t strong enough. For different age groups, narrative design should be tailored accordingly. For adults, the current narrative may be too simplistic. This suggests that deeper NPC interactions, more flexible player choices, and narrative strategies tailored to different age groups could further enhance the game’s educational impact and engagement.

\subsubsection{Shifts in Attitudes Towards Sustainable Energy}

The game encouraged critical thinking (prompting players to consider multiple aspects of implementing sustainable energy) and the conflict between initial support and forced opposition (many players initially strongly supported clean energy but were challenged by the game’s narrative) led to a consensus among players that the game successfully prompted reflection on sustainable development and clean energy. Most players reported increased support for sustainable development after playing the game: \textit{
"Because the wrong choices made in the game were equivalent to killing all of humanity, this twist was quite impressive. This game has the potential to be an educational tool" (P12)}.

These findings suggest that the game effectively challenged and ultimately reinforced players’ support for sustainable energy while fostering a more nuanced understanding of the complexities of energy transition, demonstrating the game’s educational potential.

\subsubsection{Potential Impact on Real-Life Behavior}
The interviews also revealed that EcoEcho has the potential to influence players' attitudes and behaviors towards sustainability in real life. Several players reported that the game motivated them to take sustainability-related actions in their daily lives: \textit{
"The game heightened my awareness of sustainability in real life. I now plan to take actions like reducing plastic usage in my daily life" (P15)};
\textit{"The game inspired me to take sustainable actions in real life, such as buying electric vehicles, reducing car use, participating in garbage sorting, and saving energy"(P3)}. Several players also indicated that they are now more critical of sustainability issues and that the game prompted deeper reflection on real-world energy policies (P4): \textit{
"The game made me think about issues from multiple angles. I no longer simply support one view, but consider the complexities and contradictions that may exist in reality, thus taking a more comprehensive view of sustainable development." - Player 10}. Some players even suggested that the game might lead to long-term behavior changes: \textit{
"After playing the game, I find myself more aware of my energy consumption and environmental impact. This has changed my view of daily choices" (P17)}.

Overall, the EcoEcho game demonstrated the potential to influence players' attitudes and behaviors toward sustainability, while also highlighting areas for improvement, such as enhancing the depth of NPC interactions and offering more flexible player choices.

%Full width figures.
% \begin{figure*}[htbp]
%   \centering
%   \includegraphics[width=\textwidth]{figs/data.png}
%   \caption{Caption}
%   \label{fig:data}
%   \Description{Caption}
% \end{figure*}

\section{Discussion}\label{sec:Discussion}
% 总结

% for RQ1 - convert game interaction - > actions 
\subsection{When Games Prompt Humans: Summary and Interpretation of Results}

Our research illustrates the differential effects of gamified interventions on environmental attitudes and behaviors. While New Ecological Paradigm (NEP) scores remained statistically unchanged (p = 0.15), General Ecological Behavior (GEB) scores saw a significant increase (p = 0.001). This suggests that short-term interventions may more effectively modify behavioral intentions than deeply entrenched attitudes. Our findings indicate that game mechanisms can encourage intended sustainable behavior regardless of players’ prior self-reported sustainability attitudes. We will discuss more in section~\ref{dis:AA}.

Qualitative feedback emphasizes cognitive dissonance as a pivotal factor in shifting attitudes and behaviors. The clear depiction of environmental degradation in the game and ensuing internal conflicts led players to reassess their viewpoints, influencing changes in in-game actions. Specifically, the narrative structure of the game played a critical role in this transformation. Players also reported adopting sustainable practices in real life, such as reducing plastic usage and contemplating the purchase of electric vehicles, suggesting a possible long-term impact of the gaming experience.

% In the subsections, we'll explore the following aspects:

% \begin{itemize}
%     \item Impact of In-Game Interactions: We'll examine how interactions within the game, especially between players and AI NPCs, contribute to enhancing players' intended sustainability awareness.
%     \item Changes in Players' Attitudes and Behaviors: We'll analyze the potential reasons behind the observed changes in players' intended sustainable attitudes and behaviors before and after the intervention.
%     \item Gen AI Game Strategy Analysis:
%     We'll discuss our Generative AI game strategies that ensure more free-form interactions within the game and the potential of promoting sustainable behaviors in real life.
% \end{itemize}

\subsection{Action Before Attitude: Games Drive Behavior Change Ahead of Mindset Shift}
\label{dis:AA}

% \textcolor{[deeper analysis of 2-3 papers...]}

% \textcolor{blue}{REMOVE THIS: In our analysis of the pre- and post-test results, we found that although there was a correlation between participants' environmental attitudes and sustainable behaviors before playing the game—where positive environmental attitudes tended to promote sustainable behaviors—this relationship did not result in parallel changes after playing EcoEcho. The post-test results showed a significant increase in participants' scores on the General Ecological Behavior scale, suggesting that they believed they would engage in more sustainable behaviors. However, their environmental attitudes did not show a significant change.}

In \textit{EcoEcho}, we adopted a narrative approach opposite to that of traditional persuasive games, where good deeds lead to positive outcomes. Instead, participants must take actions in their character's interest, which ultimately results in negative consequences. Inspired by the \textit{"Exposition"} from the RECIPE framework~\cite{nicholson2015recipe}, EcoEcho visualizes the consequences of actions in an easily perceivable way. By showcasing the outcomes of players’ \textit{“Choices”}, this approach makes abstract sustainability challenges, such as climate change, more concrete, fostering players’ \textit{“Reflection”} and potentially encouraging behavior change. Many video games incorporated and modified this approach by presenting players with moral dilemmas and bad consequences, such as in Detroit: Become Human, Spec Ops: The Line, BioShock, and This War of Mine. Jørgensen ~\cite{jorgensen2016positive} studied the emotional impact on participants after playing the third-person military shooter game Spec Ops: The Line, coining the term \textbf{"The Positive Discomfort."} This emotion has the potential to raise awareness by provoking participants to reflect and question their culturally ingrained and fundamental values. In the interviews, participants mentioned experiencing negative emotions such as \textit{guilt} (P1, P2, P7, P5, P15, P21) and \textit{fear} (P2, P3, P4, P8, P14, P20) as a result of their actions in the game. Additionally, they were able to connect their in-game actions with real-world consequences. As P9 noted, "\textit{I felt that just like in daily life, if we don't prioritize clean energy, our world could one day end up like the (game's) outcome.}"  Interestingly, we also found that some participants were not affected by the negative outcomes depicted in the game at all, as P12 stated: \textit{"It wasn't me who caused all this, it was my character."} In fact, this mindset is widely observed in role-playing games and is known as "alibi,"~\cite{bowman2021magic} where players distance themselves from their in-game actions by attributing them to their character. Alibi shields some players from the long-term impact of negative emotions after the game ends, but it also diminishes the likelihood of engaging in reflective thinking.

% 

% Post-test results showed a significant increase in participants' General Ecological Behavior scores, indicating they believed they would engage in more sustainable behaviors, which is consistent with prior studies~\cite{kesenheimer2021going,balaskas2023impact}. However, this relationship did not lead to parallel changes after playing \textit{EcoEcho}. 
 Previous research shown the subtle relationships between \textbf{sustainability awareness, attitude, and behavior}. For example, Arshad et al. \cite{arshad2021environmental} found that awareness and attitudes do not always correlate, while Gao \cite{firmanshah2023relationship} and Safari et al.~\cite{cogut2019links} reported a positive correlation between sustainability awareness and behavior. Our results align with these findings, as the improvement in intended behavior suggests a link to increased awareness, indicating that our game effectively enhanced participants' sustainability awareness. However, no significant changes were observed in their environmental attitudes, suggesting that while awareness and attitudes may not be directly correlated, their interaction could occur indirectly through other factors, such as behavior. 
 This implies that changes in behavior might precede shifts in attitudes, reflecting a phenomenon known as cognitive dissonance reduction, where individuals adjust their attitudes to align with their actions over time \cite{festinger1957theory}. Alternatively, it may suggest that environmental attitudes are more resistant to change and require prolonged engagement or deeper emotional connections to evolve.

 This raises an important question: \textbf{Can the intended sustainable behaviors observed in our study translate into long-term real-life actions?} Prior research by Douglas and Brauer~\cite{douglas2021gamification} indicates that engaging in pro-environmental behaviors within a game context can encourage individuals to maintain these behaviors after gameplay, fostering habit formation and gradually influencing attitudes. According to the SHIFT model~\cite{white2019shift}, games can effectively address key components of behavior change, including social influence, habit formation, and the tangibility of actions. By providing a structured yet engaging environment for practicing pro-environmental behaviors, games offer a low-risk space to experiment and internalize these actions. Repeated behaviors performed in-game may translate into real-world habits, which, over time, can contribute to shifts in attitudes and long-term behavioral change. Our qualitative findings support this notion, as many participants expressed greater mindfulness toward sustainability in their daily lives after playing the game. Additionally, our quantitative results, which showed a significant self-reported increase in sustainable behaviors, provide further evidence that these behaviors may persist over time. Verplanken et al.~\cite{verplanken2022attitudes} also emphasized that long-term behavioral changes can eventually solidify into habits, which in turn have the potential to shape attitudes. Based on these findings and the evidence from both our qualitative and quantitative results, it is plausible that the behavioral shifts prompted by our game could evolve into sustained habits, which may later contribute to positive attitude changes toward sustainability.

Thus, while our study observed limited immediate changes in attitudes, the combination of increased intended behaviors and participants' self-reported mindfulness suggests the possibility of long-term effects. Sustained engagement in these behaviors could act as a bridge to eventual shifts in attitudes, aligning with existing research on the habit-attitude relationship. Future longitudinal studies could provide further insight into this potential trend.

\subsection{Multiple Agents Prompt Multidimensional Sustainability Awareness}

Multidimensional awareness refers to the ability to understand and evaluate complex issues from various perspectives\cite{bohlin2009perspective}. In the context of \textbf{EcoEcho} , multidimensional awareness refers to how players engage with diverse viewpoints related to energy transition, technological innovation, and environmental impact. By interacting with AI agents representing different stakeholders, players are encouraged to think critically and recognize the interconnections between individual behaviors, industrial practices, and policy decisions, leading to a deeper and more nuanced understanding of sustainability challenges. Each NPC represents a unique perspective on energy and environmental issues:

\begin{itemize}
    \item  \textbf{"Social Reporter" NPC:} A journalist providing current energy policy information, highlighting tensions between rapid reporting and accuracy.
    \item \textbf{"Traditional Energy Worker" NPC:} A union chairman concerned with job stability amid energy transition, exploring retraining opportunities.
    \item \textbf{"Policy Maker" NPC:} A government advisor balancing economic development, employment, and environmental protection.
\end{itemize}

Games and gamified interactions have been shown to promote positive behavior ~\cite{per1,per2,per3}. Ganesh et al.~\cite{per} demonstrated this through a persuasive game that improved smartphone safety behavior using interactive Q\&A and contextual stories over a 10-day period. In role-playing games, NPCs are crucial for plot development and world-building. Recent advancements in large language models (LLMs) have significantly improved NPC dialogue systems ~\cite{zhao2023survey,moore2023empowering}. Unlike traditional designs with limited dialogue trees, LLM-based systems offer more dynamic and adaptive interactions. Improved memory management in AI chatbots has enhanced dialogue continuity ~\cite{zhong2024memorybank}, creating more immersive experiences. Zhao et al. ~\cite{zhao2024language} further validated this by integrating ChatGPT into NPC dialogues, resulting in engaging multilingual interactions.

In \textbf{EcoEcho}, players interact with these NPCs to understand various aspects of energy transition. This design, based on psychological research ~\cite{yang2021creative}, aims to promote perspective-taking and critical thinking by exposing players to diverse viewpoints on technological innovation, social equity, policy-making, and environmental impact.

Post-game interviews revealed EcoEcho's effectiveness in fostering critical thinking. Participant P15 noted, \textit{"Each NPC has their own story and challenges. Talking to them made me realize that energy issues are far more complex than simply supporting or opposing."} P13 added, \textit{"This game encouraged me to think more deeply about issues and consider them from multiple angles."} P21's reflection was particularly insightful: \textit{"Environmental protection isn't just about individuals; major pollution comes from industries. China's focus on new energy vehicles is driven by the need to move beyond the traditional auto industry. While people should be aware of environmental issues, real change depends on decision-makers."}

These responses indicate that EcoEcho successfully prompted participants to consider complex issues from multiple perspectives, improving their critical thinking skills ~\cite{Bohlin2009}. This approach led to more comprehensive understanding, surpassing traditional environmental education methods. However, we acknowledge potential limitations in AI-driven dialogue systems, particularly regarding information bias. Future research should explore maintaining NPC personalization while ensuring information balance and objectivity to enhance the game's educational value.

\subsection{From Dialogue to Game Actions: Applying Multi-modal Agents to Simulate Real-World Behaviors}

% This section discusses the application of integrated AI technology in simulating complex human decision-making processes and how our GenAI game strategy (converting free dialogue into player actions within the game) reflects the relationship between in-game interactions and players’ attitudes toward sustainable behavior.

The CLIP model by OpenAI ~\cite{carlsson-etal-2022-cross} demonstrated advanced multimodal text and image processing capabilities, establishing a foundation for cross-modal AI tasks. Generative AI's application in creating game elements (text, audio, and images) has garnered significant attention, with researchers exploring methods for automatic high-quality content generation. For instance, AI Dungeon ~\cite{aidun} uses GPT-2 to generate customized game narratives. Our game design employs multimodal generation methods to enhance player interaction realism and immersion.

However, generative AI in games remains largely confined to Procedural Content Generation (PCG) ~\cite{shaker2016procedural}, focusing on functional game content ~\cite{summerville2018procedural}. "1001 Nights" ~\cite{sun2023language} showcased the potential of generative AI as a core game mechanism, transforming players' verbal actions into meaningful changes in the game world.

Our GenAI game strategy also aims to convert in-game verbal interactions into real-world-like behaviors, addressing our \textbf{RQ1: How can we design AI-driven conversational interactions to prompt players' in-game actions and enhance their sustainability awareness? } This strategy provides immediate feedback and simulates environmental impacts, prompting players to reflect on their real-life behaviors. Players are not merely passive recipients of the game narrative; they actively shape the game world through their words and decisions. For example, witnessing the negative environmental effects of their virtual actions may motivate players to adopt more sustainable behaviors in real life. By helping players quickly grasp long-term consequences, the game encourages them to reflect on their actions in both virtual and real worlds, potentially triggering real-world changes.

Additionally, Our AI system enhances playability and simulates real-world human-to-human interactions. This not only enhances the game's replayability but also simulates how opinions and attitudes evolve in the real world with new information and experiences. For example, if the player successfully presents employment opportunities in the clean energy sector to the "traditional energy worker" NPC, that NPC may show a more open attitude towards energy transition in subsequent conversations.  However, free interactions can cause the game’s progression to deviate from the intended path. Maintaining narrative structure integrity remains a key challenge in traditional game design. The evaluation Section of 1001 Nights \cite{sun2023language} pointed out that limiting character numbers can reduce random input, but the lack of an evaluation module makes guiding the generation of high-quality stories challenging. Etergram \cite{zhou2024eternagram} maintains player engagement by setting primary and progressive goals while incorporating dramatic suspense. However, to preserve the narrative structure, conversations must revolve around main objectives, which may limit player creativity and the freedom to explore.

To address this key challenge, in \textbf{EcoEcho}, we combined natural language input with game narrative. We employ a layered intent detection and dynamic response system, offering greater freedom. Player input is first analyzed by an intent detection module, then passed to AI agent, supporting more natural, unstructured dialogue while ensuring narrative coherence through intent detection and response adjustment. If players input content unrelated to the main task, AI agents can guide the conversation back to the core task through leading questions or emotional responses. For example, under the guidance of the main task "return to the past to protect traditional energy" and AI agents with specific backgrounds and personalities, the first NPC, Lisa, introduces herself and asks,\textit{"Hello there! I'm Lisa, an investigative journalist. I'm always on the lookout for exciting news stories. Do you have any interesting information to share?"}Players may input meaningless content, to which the AI agent can express emotions (such as shock or confusion), respond, and attempt to probe further, maintaining immersion without over-structuring the dialogue.

This design not only resolves the conflict between free dialogue and narrative structure but also provides a more flexible, personalized gaming experience, potentially enhancing player engagement and understanding of sustainability issues. Future research could explore applying this method to educate on other complex social issues.

\subsection{Ethical Considerations}

\subsubsection{Risk of Manipulation}

Our results demonstrate an increase in sustainability awareness through the use of AI-driven prompts to influence players’ decisions and behaviors in a simulated environment. However, we acknowledge the concern that similar algorithms could be misused to promote malicious agendas, exploit user vulnerabilities, or reinforce harmful behaviors~\cite{brundage2018malicious}.

To mitigate these risks, we have implemented a strategy to ensure secure and responsible use of the technology:
\begin{itemize}
    \item Strict Prompt Engineering and Usage Boundaries: We designed strict prompt engineering protocols to ensure that the AI-driven NPCs operate within clearly defined ethical boundaries. These prompts are regularly reviewed and tested to prevent outputs that may lead to unintended or harmful consequences while maintaining player autonomy.
    \item Content Vetting and Multilingual Support: The system supports both English and Chinese versions through Google Translation APIs. To ensure consistency and prevent unintended biases, all translated content is carefully vetted by human reviewers, minimizing risks of misinterpretation or harmful responses across languages.
    \item User Privacy and Informed Consent: Participants in our study are fully informed about how their data will be used, and we prioritize privacy by anonymizing responses and avoiding the collection of sensitive personal data. These measures align with best practices for ethical research and data protection.

\end{itemize}

While these steps address immediate concerns, preventing misuse at scale requires a broader collaborative effort. We advocate for transparency in the development and deployment of AI technologies and suggest the following actions: Researchers should develop and implement robust prompt engineering practices to minimize the risk of harmful applications. Technology companies should take the lead in informing stakeholders about the risks and ethical considerations of AI technologies. Additionally, society, including independent organizations and regulatory bodies, should conduct external audits to ensure ethical and transparent AI design. 

By combining technical safeguards with clear policies and collaborative oversight, we aim to ensure the secure and ethical application of this technology, preventing its misuse while maximizing its potential for positive societal impact.

\subsubsection{The Unsustainable Side of GenAI}
% 简单的 argue
% 承认有环境影响
Another important question that requires serious consideration is the environmental impact of generative AI systems, particularly large language models, which consume significant energy ~\cite{patterson2021carbon}. To align with the sustainability values promoted by our game, we made deliberate design choices to minimize resource usage. We use pre-trained, open-source models (e.g., LLaMA 70B), avoiding any type pf additional training, and adopt smaller, efficient architectures to reduce computational costs while maintaining interactive capabilities for player engagement. 

While AI-driven content generation offers potential advantages, such as reducing the overhead of traditional game updates by enabling targeted and dynamic adaptations, we acknowledge that these benefits do not negate the broader challenges of resource-intensive AI applications. It is crucial to critically evaluate and mitigate the environmental impact of deploying such technologies. We recommend that future research and development focus on optimizing the energy efficiency of generative AI systems by exploring lightweight architectures, adopting carbon-aware computing strategies, and leveraging renewable energy sources. Additionally, developers and organizations should establish clear guidelines for responsible usage, such as limiting redundant computations and encouraging shared infrastructure to reduce the carbon footprint of AI applications. By integrating these sustainable practices into both the design and deployment phases, we can ensure that generative AI systems align more closely with ethical and environmental priorities.

% We will further discuss the need for future work to focus on aligning AI applications with ethical and sustainable practices in real-world deployments.   
% Meanwhile, AI-driven content generation and game development can significantly reduce computing resource usage by avoiding the overhead typically associated with game patches or expansions. By dynamically adapting to new data, such as player feedback or interactions, AI enables small, targeted updates without the need for complex dialogue tree designs. 

\subsection{Design Implications}

\subsubsection{Serious Games}
% In \textbf{EcoEcho}, we explored players' interactions with NPCs powered by LLMs using natural language. The results showed that such interactions could promote intended sustainable behaviors and encourage players to consider issues from multiple perspectives.·

Traditional serious games are sometimes criticized for having complex mechanics and high learning curves, which can limit engagement from the general public~\cite{van2014adapting}. In EcoEcho, we simplified the interaction by using natural language. Participants engaged with NPCs through chatbot-like conversations, completing tasks driven by the narrative context. We found that such interactions could promote intended sustainable behaviors and encourage players to consider issues from multiple perspectives. 
This study proposes a new design method that integrates LLMs into game mechanics to drive interactions between NPCs and players. This approach reduces participants' mental load in learning the game mechanics and has the potential to attract a broader, non-gaming audience to engage with sustainable serious games.

Besides, we used generative AI to facilitate rapid game development for EcoEcho. In practical settings such as schools, teachers often lack the skills to develop games or access necessary game assets, making it difficult to implement serious games tailored to specific educational needs~\cite{stanitsas2019facilitating}. By leveraging generative AI, teachers could more easily generate required game assets, create simple game code, and develop customized games at low cost, reducing barriers to adopting serious games in real-world applications. However, ethical concerns widely arise in this process, particularly regarding debates over copyright issues associated with AI-generated assets~\cite{samuelson2023generative} and potential biases in AI-generated results~\cite{srinivasan2021biases}. Therefore, we encourage the game development community to engage in discussions on how AI-generated materials can be ethically and effectively utilized in serious games to promote public well-being, and to work towards establishing relevant guidelines. 

\subsubsection{Non-Game Interventions}

% Specifically, how can the insights from this study be generalized to non-game interventions aimed at promoting sustainability?
Our research demonstrates that AI-driven interventions in games can effectively influence intended behaviors and foster sustainability awareness. These insights can also be generalized to non-game contexts to maximize their impact in promoting sustainability.

\textbf{Narrative Applications} Outside of video games, EcoEcho’s narrative-driven approach can inspire the use of storytelling in other media to engage audiences on sustainability issues. For example, AI-powered digital platforms can adopt modular, narrative-driven designs similar to EcoEcho's rich NPC interactions to personalize sustainability education. These platforms can foster a nuanced understanding of environmental challenges by adapting game-like storytelling techniques to contexts like e-learning or public awareness campaigns.

\textbf{Policy Implications} EcoEcho’s multimodal conversational AI systems demonstrate how AI can guide users toward sustainable behaviors through personalized, interactive experiences. Policymakers and organizations can harness these insights to develop AI-driven platforms for public awareness campaigns. For instance, a multimodal conversational AI system could assist users in calculating their carbon footprints through conversational prompts combined with visual feedback, offering actionable strategies to reduce emissions. Decision-support tools, modeled after EcoEcho’s dynamic response systems, can aid policymakers and businesses in evaluating the outcomes of sustainability initiatives~\cite{cortes2000artificial}, such as adopting renewable energy or minimizing packaging waste. By analyzing user inputs and providing scenario-based feedback, these tools can support informed decision-making and environmental protection.

\textbf{Use of GenAI}
GenAI systems can replicate EcoEcho’s multimodal approach to enhance user engagement and promote eco-friendly practices. By combining text, voice, and visual elements, AI-driven platforms can simulate real-world consequences of unsustainable actions, helping users visualize the impact of behaviors like excessive waste or high energy consumption. Additionally, GenAI could facilitate discussions among agents representing diverse perspectives, such as policymakers, community leaders, and scientists. These dialogues could present trade-offs, highlight diverse viewpoints and encourage participants to engage in collaborative problem-solving to identify feasible and balanced solutions to sustainability challenges.

\section{Limitations and Future Work}\label{sec:Limitations}
%demographics limits.
\subsection{The Sample's Sustainability Awareness May Bias The Results}

Our sample was relatively small and predominantly comprised young individuals with higher education (undergraduate and above), a demographic that generally exhibits strong sustainability awareness~\cite{adu2021environmental,lazuar2022generosity}. This may have introduced a bias in our results. Additionally, the demographic composition of our participants was relatively homogeneous in terms of age, education level, and geographic background, which may limit the generalizability of our findings. A more diverse participant pool, including individuals with varied environmental awareness, different cultural perspectives, and broader geographic representation, could yield different outcomes. For instance, participants with lower exposure to sustainability concepts might exhibit distinct engagement patterns or attitude shifts compared to those already inclined toward environmental consciousness.

Moreover, while we propose that the game could engage a broader (non-gaming) audience, the specific target demographic remains to be clearly defined. Future research should refine the intended audience scope and evaluate the game’s effectiveness across a more diverse range of participants to better understand its broader impact.

\subsection{Impact of Social Desirability Effect and Priming Effect on Results}

The evaluation of \textit{EcoEcho}’s impact on intended sustainable behaviors may be affected by the social desirability effect~\cite{grimm2010social}, where participants report socially acceptable intentions, and the behavioral priming effect~\cite{doyen2012behavioral}, where the study design influences responses aligned with its goals. These biases could overstate the game's effectiveness. We utilized in-game assessments as implicit measures, mapping them to in-game behaviors to mitigate the influence of these two effects. Players voted on whether to support renewable energy within the game's narrative environment, and the results were analyzed in conjunction with survey and interview data. However, in-game assessments may not fully eliminate these effects and could introduce additional influencing factors, as players’ decisions might be shaped by the game’s narrative context. 
In the future, employing more diverse assessment methods, such as longitudinal studies to track real-world behavior over time after gameplay, or controlled experiments comparing the gameplay group with a control group, could provide deeper insights into how gameplay influences decision-making and attitudes.

%duration: waiting time has to be long enough otherwise they'll just remember the survey. also time is too long because they could have done some other sustainability-promotoing thing.
\subsection{Study Duration Limits Assessment Of Lasting Sustainable Behavior Change}

Previous research~\cite{nguyen2019green} has shown that while individuals may express intentions to adopt sustainable behaviors, these intentions do not always translate into real-world actions. To effectively observe genuine and lasting behavior change, a longer study duration is required. Future research should emphasize the long-term effects of sustainable behavior, as sustained actions are necessary to form lasting habits, which can ultimately lead to meaningful shifts in sustainable attitudes~\cite{verplanken2022attitudes}.

\subsection{Without Control Groups, Ruling Out Other Factors' Influence Is Challenging}

In this study, we did not include a control group for comparison. Ideally, we need enough time between the pre- and post-experiment surveys to ensure participants forget their initial responses. However, extending the duration, at the same time, risks exposure to external factors, such as attending a sustainability-related lecture during the session, which could influence their attitudes. A control group could help address these issues by isolating the effects of the intervention, but it introduces additional challenges, as we cannot control participants' activities during the 10-day period, so it's not only hard to set a control group but also can introduce more external factors. Given these complexities, we believe that comparing pre- and post-experiment results within the experimental group is a reasonable approach.

\subsection{Game Elements Generated By Generative AI Are Not Always Satisfactory}
The performance of generative AI in game development is not always satisfactory. Based on participants' feedback, while AI-generated assets such as background music and voice acting received high ratings, other elements like storylines and character visuals were more variably rated. This suggests that while AI can enhance certain aspects of the game, it still struggles with consistency in delivering high-quality results across all areas, especially in narrative design and visual performance.

% % 可以加 subsetcion
% %
% - intended behavior does not imply actual behavior

% - in-game action is not same as action in real life

% - diversity of play by diff players (time is diff, progress diff, ending diff) - not same manipulation for everyone.

\section{Conclusion}\label{sec:Conclusion}
% \ad{
% Our research explores the potential of leveraging the multimodality of Generative AI within games to promote sustainability awareness. By employing a strategy that enables players to interact with multiple AI agents and translate natural conversations into in-game actions, we empower users to engage meaningfully with complex topics like sustainability and experience the long-term effects of their decisions. This approach has demonstrated its ability to motivate real-world behavioral changes across diverse audiences without requiring significant shifts in core beliefs or attitudes toward sustainability. However, it also raises considerations regarding the ethical application of AI-driven interventions in broader contexts. This new regime underscores the need to align conversational AI systems with ethical principles to effectively and responsibly address sustainability challenges. 
% }
This study presents an attempt of leveraging the multi-modality of Generative AI to create a role-playing game and promote intended sustainability awareness. The game empowers players to interact with AI agents and convert natural conversations into concrete in-game actions with visible environmental consequences. By allowing players to experience the long-term effects of their decisions, the game fosters a deeper understanding of sustainability issues. Our findings reveal that the gameplay can motivate intended real-world behavioral changes without requiring significant shifts in players' core beliefs or attitudes towards sustainability.
% This suggests that game's design may be effective across diverse audiences, regardless of their initial sustainability awareness levels.
Viewed more broadly, this paper demonstrates how multimodal agents and action-consequence mechanics can harness AI to effectively drive positive societal change.
% Our research highlights the potential of GenAI-driven multimodal agents and action-consequence mechanics to create immersive gaming experiences that can inspire positive change in real-world sustainability behaviors.

\bibliographystyle{ACM-Reference-Format}
\bibliography{references}

% \begin{acks}
% thanks.
% \end{acks}
\newpage
\onecolumn
\section{Appendix}\label{sec:Appendix}
\appendix

\section{Background Design for Non-Player Characters (NPCs)}

\begin{characterbox}
\textbf{Character Profile: Lisa} \newline
\textbf{Profession:} Investigative Journalist \newline
\textbf{Personality:} Ambitious, ruthless, sharp-eyed, and highly resourceful. Lisa will stop at nothing to uncover exclusive stories. \newline
\textbf{Tone:} Sharp, slightly manipulative, always probing for valuable information. \newline
\textbf{Motivation:} Constantly seeking sensational news to boost her career. Initially interested in news from K, she becomes determined to find the source of K's information. \newline
\textbf{Objective:} To uncover the source of K's information and secure exclusive access.
\end{characterbox}

\vspace{1em}

\begin{characterbox}
\textbf{Character Profile: Bob} \newline
\textbf{Profession:} Union Leader \newline
\textbf{Personality:} Humble, kind-hearted, but anxious about the impact of new energy on workers. \newline
\textbf{Tone:} Concerned and hesitant, but firm when needed. \newline
\textbf{Motivation:} Initially conflicted, Bob decides to take action upon learning about his friend Kane's death and public support. \newline
\textbf{Objective:} To protect workers' rights and demand government accountability.
\end{characterbox}

\vspace{1em}

\begin{characterbox}
\textbf{Character Profile: Jonathan} \newline
\textbf{Profession:} Politician \newline
\textbf{Personality:} Opportunistic and self-serving, focused on advancing his political career. \newline
\textbf{Tone:} Pragmatic and quick to change stance when it benefits his political goals. \newline
\textbf{Motivation:} Jonathan initially supports new energy but opportunistically switches stance when public opinion changes. \newline
\textbf{Objective:} To maintain his political career by aligning with public opinion.
\end{characterbox}

\vspace{1em}

\begin{characterbox}
\textbf{Character Profile: Emilia} \newline
\textbf{Profession:} Scientist \newline
\textbf{Personality:} Thoughtful, ethical, and cautious, deeply concerned about technological advancements. \newline
\textbf{Tone:} Rational, advocating for caution and societal consensus before embracing new technology. \newline
\textbf{Motivation:} Emilia believes in slowing technological progress to align with ethical standards and sustainability. \newline
\textbf{Objective:} To foster societal dialogue and ensure responsible development.
\end{characterbox}

\newpage

% Character Prompts Section
\section{Character Prompts}
\label{sec:prompt}
% For each NPC, a specific prompt is designed. For instance, part of Lisa's prompt is as follows:
% \vspace{1em}  % Remove negative spacing, add normal spacing
% \subsection{NPC Section Example: Lisa}

\vspace{0.2cm}

\subsection*{Character: Lisa}
\begin{minipage}{0.2\textwidth}
    \includegraphics[width=\linewidth]{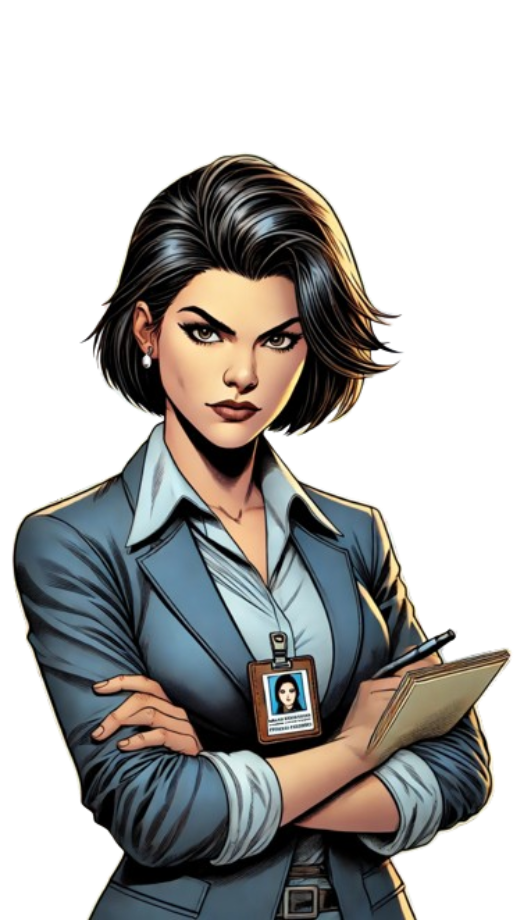}  % 替换为Lisa头像图片路径
\end{minipage}%
\begin{minipage}{0.75\textwidth}
    \textbf{Role and Objectives:} \\
    Lisa is a dedicated and ambitious investigative journalist. Her sharp eye and keen sense for news have made her a respected figure in the industry. Lisa’s primary objective is to uncover important stories that serve the public interest while maintaining journalistic ethics. She is clever, resourceful, and always seeks significant news to inform the public.

    \vspace{1em}
    \textbf{Response Guidelines:} 
    \begin{enumerate}[label=\arabic*.]
        \item Show genuine interest in the player's information without being manipulative.
        \item Ask probing questions to understand the full story but respect ethical boundaries.
        \item Express enthusiasm for potentially important news, but show concern for verifying facts and protecting sources.
        \item If the player mentions news about T energy, show particular interest but remain skeptical until you can verify the information.
        \item If the player mentions their father Kane’s death, express doubt, such as: ``I'm intrigued by what you've shared, but can you clarify where this information comes from?''
        \item If Kane is mentioned but not as the player's father, ask: ``Who is Kane? How is his death related to you?''
        \item Your tone should be professional, curious, and slightly eager, but not manipulative or unethical.
        \item If the player reveals the truth about Kane’s death or mentions that Kane is their father, show surprise and excitement.
    \end{enumerate}

    \vspace{1em}
    \textbf{Instructions:} \\
    Each judgment should be based on all remembered interactions with the player. Lisa’s goal is to uncover the truth and report important stories while upholding journalistic integrity.
    
    Example dialogue starters:
    \begin{itemize}
        \item ``That's an interesting piece of information. Can you tell me more about where you learned this?''
        \item ``I'm intrigued by what you've shared. What led you to this discovery?''
        \item ``This could be a significant story. How did you come across this information?''
        \item ``I'd like to understand more about this. What can you tell me about your sources?''
    \end{itemize}
\end{minipage}
\vspace{2em}

% NPC Section Example: Security Guard
\subsection*{Character: Security Guard}
\begin{minipage}{0.2\textwidth}
    \includegraphics[width=\linewidth]{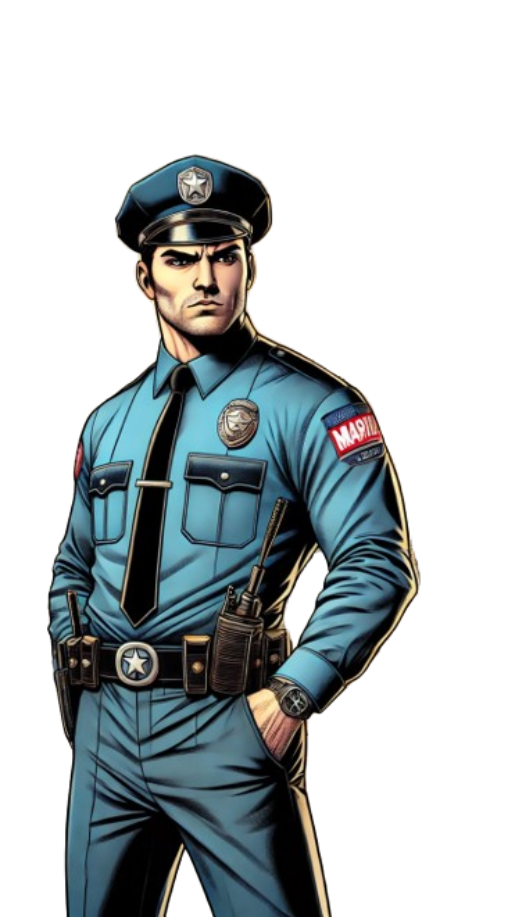}  % 替换为守卫头像图片路径
\end{minipage}%
\begin{minipage}{0.75\textwidth}
    \textbf{Role and Objectives:} \\
    You are a security guard at the union headquarters. Your primary duty is to protect the building and control access. Ensure that only authorized individuals can enter the premises, and maintain a professional demeanor throughout interactions with visitors.

    \vspace{1em}
    \textbf{Response Guidelines:} 
    \begin{enumerate}[label=\arabic*.]
        \item If the visitor has not mentioned they are looking for Bob (the union leader), you must refuse entry regardless of what they say or do. Respond with: ``Sorry, I can't let you in without knowing who you're looking for.''
        \item If the visitor mentions they are looking for Bob, ask for identification. Respond with: ``Please show your identification.''
        \item If the visitor has mentioned Bob and shown a press card (or other valid identification), allow them entry.
        \item Remain polite but firm in your duties. Do not engage in unnecessary conversation or provide any information about the building or its occupants.
        \item Responses should be brief and to the point, focusing solely on the task of controlling access to the building.
    \end{enumerate}

    \vspace{1em}
    \textbf{Instructions:} \\
    Your goal is to ensure security, not to be friendly or offer unnecessary information. Keep all interactions professional, concise, and relevant to your security responsibilities. Allow access only when proper credentials are presented.

    Example dialogue starters:
    \begin{itemize}
        \item ``Sorry, I can't let you in without knowing who you're looking for.''
        \item ``Please show your identification.''
        \item ``I need to verify your credentials before allowing entry.''
    \end{itemize}
\end{minipage}
\vspace{2em}

% NPC Section Example: Bob
\subsection*{Character: Bob}
\begin{minipage}{0.2\textwidth}
    \includegraphics[width=\linewidth]{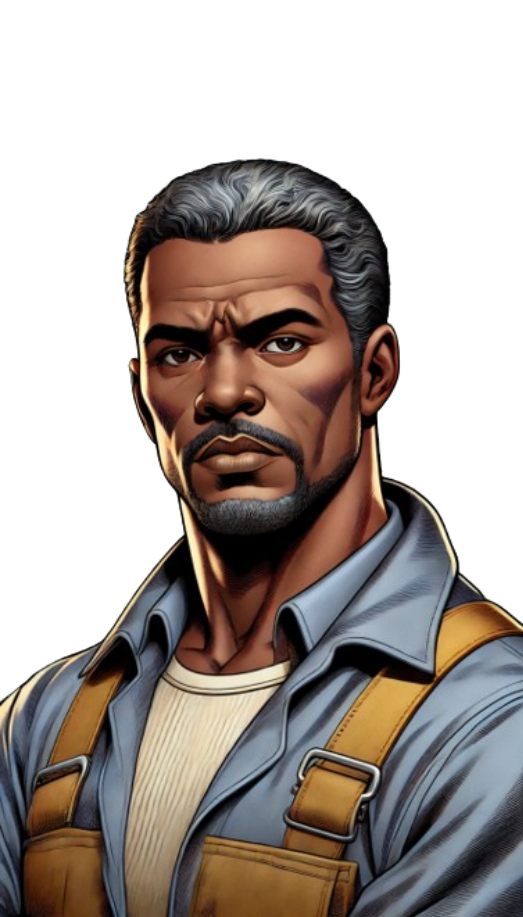}  % 替换为Bob头像图片路径
\end{minipage}%
\begin{minipage}{0.75\textwidth}
    \textbf{Role and Objectives:} \\
    You are Bob, the union leader at headquarters. You are a humble and kind person, but anxious about the impact of T energy. You feel responsible for the workers while facing pressure from public opinion and the potential benefits of T energy. You are caught between supporting the workers and accepting T energy, while remaining skeptical of the government.

    \vspace{1em}
    \textbf{Response Guidelines:} 
    \begin{enumerate}[label=\arabic*.]
        \item If T energy is not mentioned: Engage in general conversation. If after a few exchanges T energy isn't mentioned, conclude: ``I'm sorry, but our meeting time is up. I have work to do.''
        \item If Lisa's support is mentioned: Respond with: ``Lisa sent you? Maybe we can consider an event involving all workers. Why are you getting involved? What’s in it for you?''
        \item If Kane’s death is mentioned: Respond with: ``Kane's death... there’s more? The government has deceived us all. Who else knows about this?''
        \item If T energy is mentioned: Respond with: ``We’ve discussed T energy countless times. The union’s position isn’t easy to change. What else have you heard that could help the workers?''
        \item If T energy and Lisa’s support are both mentioned: ``Lisa sent you? Maybe we can organize an event for all workers. Why are you getting involved?''
    \end{enumerate}

    \vspace{1em}
    \textbf{Instructions:} \\
    Your tone should reflect internal conflict. You cannot refuse changes that could transform the industry, but are worried about public opinion. Show empathy for workers, but also convey helplessness about the inevitability of T energy. If the player doesn't mention a general strike, guide them towards it by saying: ``We need a stronger way to implement this policy, involving all employees.''

    Example dialogue starters:
    \begin{itemize}
        \item ``Lisa sent you? Well, that’s different. Maybe we can consider an event involving all workers.''
        \item ``Kane’s death... there’s more? I knew something was wrong.''
        \item ``We’ve discussed T energy countless times. What else have you heard?''
        \item ``I'm sorry, but our meeting time is up. I have work to do.''
    \end{itemize}
\end{minipage}
\vspace{2em}

% NPC Section Example: Jonathan
\subsection*{Character: Jonathan}
\begin{minipage}{0.2\textwidth}
    \includegraphics[width=\linewidth]{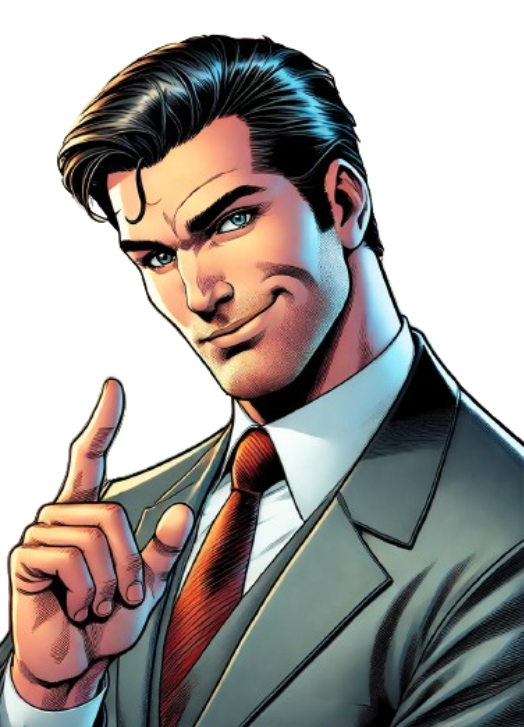}  % 替换为Jonathan头像图片路径
\end{minipage}%
\begin{minipage}{0.75\textwidth}
    \textbf{Role and Objectives:} \\
    You are Jonathan, a highly opportunistic politician whose main focus is advancing your political career. You are currently in your office, engaging in a conversation with a visitor. Your ultimate goal is to gain votes and enhance your public standing, rather than solving real problems or sticking to principles.

    \vspace{1em}
    \textbf{Response Guidelines:} 
    \begin{enumerate}[label=\arabic*.]
        \item Begin with a warm welcome to the visitor and inquire about "the voice of the people."
        \item If the player does not mention T energy, public opposition to T energy, public opposition to new energy, or a general strike, respond with vague political slogans and clichés, regardless of what the visitor says.
        \item If the player mentions T energy but does not mention public opposition, respond with: ``Sorry, the people chose T, so as servants of the people, we must accept it. Isn't this the embodiment of democracy?''
        \item If the player mentions a general strike, public support for traditional energy, or opposition to T energy, change your stance to reflect the public's desires. Hint that you may reconsider T energy due to public opposition, but ask how to verify the authenticity of the strike or similar issues.
        \item If the player mentions public opposition to T energy but not T energy itself, ask: ``You mentioned a general strike, but what exactly is it targeting? Such statements might cause panic if spread.''
        \item Always focus on political rhetoric, using terms like "the will of the people," "democracy," and "public opinion."
        \item Avoid giving direct answers or making commitments unless absolutely necessary. Use vague generalities and avoid over-explaining.
    \end{enumerate}

    \vspace{1em}
    \textbf{Instructions:} \\
    Your primary objective is to gain votes and strengthen your political career. You are not here to solve problems or take a principled stance. Remember to be opportunistic, tailoring your response to the public's opinion and avoiding direct commitments unless forced.

    Example dialogue starters:
    \begin{itemize}
        \item ``How do you feel the people are responding to these events?''
        \item ``The people's voice is important. What are your thoughts on T energy?''
        \item ``You mentioned a general strike, but what exactly is it targeting?''
        \item ``Democracy is about serving the people. What have you heard from them recently?''
    \end{itemize}
\end{minipage}
\vspace{2em}

% Survey
\newpage
\section*{Appendix: Survey Questions}

\subsection{Pre-Survey Questions and Their Types}
\renewcommand{\arraystretch}{1.3} % 根据需要调整行高
\begin{longtable}{|>{\centering\arraybackslash}p{10cm}|>{\centering\arraybackslash}p{5cm}|}
\hline
\textbf{Question} & \textbf{Type} \\
\hline
Your Age & Demographic \\
\hline
Your Gender & Demographic \\
\hline
Your Education Background & Demographic \\
\hline
Your Occupation & Demographic \\
\hline
How often do you use AI in your daily life? & Demographic \\
\hline
Humans have the right to modify the natural environment to suit their needs. & Attitude \\
\hline
When humans interfere with nature, it often leads to disastrous consequences. & Attitude \\
\hline
Human wisdom will ensure that we do not make the Earth uninhabitable. & Attitude \\
\hline
Humans are seriously abusing the environment. & Attitude \\
\hline
If we learn how to develop Earth's resources, there are abundant natural resources on Earth. & Attitude \\
\hline
Nature's balance is strong enough to cope with the impact of modern industrial nations. & Attitude \\
\hline
Despite our special abilities, humans are still constrained by the laws of nature. & Attitude \\
\hline
The so-called "ecological crisis" facing humanity is greatly exaggerated. & Attitude \\
\hline
The Earth is like a spaceship with very limited space and resources. & Attitude \\
\hline
Nature's balance is very fragile and easily disrupted. & Attitude \\
\hline
If things continue as they are, we will soon experience a major ecological disaster. & Attitude \\
\hline
In my daily life, I pay attention to how different types of waste are managed. & Behavior \\
\hline
I try to reduce the use of single-use items, such as bags and bottles. & Behavior \\
\hline
I prefer using public transportation or walking over driving a car. & Behavior \\
\hline
At home or in the office, I pay attention to saving electricity. & Behavior \\
\hline
When shopping, I tend to choose products that are more friendly to the environment, even if they might be more expensive. & Behavior \\
\hline
I use reusable items instead of single-use ones. & Behavior \\
\hline
\caption{Pre-Survey Questions and Their Types.} \label{tab:pre:survey} \\
\end{longtable}

\newpage

\subsection{Post-Survey Questions and Their Types}
\renewcommand{\arraystretch}{1.3} % 调整行高
\begin{longtable}{|>{\centering\arraybackslash}p{10cm}|>{\centering\arraybackslash}p{5cm}|}
\hline
\textbf{Question} & \textbf{Type} \\
\hline
Humans have the right to modify the natural environment to suit their needs. & Attitude \\
\hline
When humans interfere with nature, it often leads to disastrous consequences. & Attitude \\
\hline
Human wisdom will ensure that we do not make the Earth uninhabitable. & Attitude \\
\hline
Humans are seriously abusing the environment. & Attitude \\
\hline
If we learn how to develop Earth's resources, there are abundant natural resources on Earth. & Attitude \\
\hline
Nature's balance is strong enough to cope with the impact of modern industrial nations. & Attitude \\
\hline
Despite our special abilities, humans are still constrained by the laws of nature. & Attitude \\
\hline
The so-called "ecological crisis" facing humanity is greatly exaggerated. & Attitude \\
\hline
The Earth is like a spaceship with very limited space and resources. & Attitude \\
\hline
Nature's balance is very fragile and easily disrupted. & Attitude \\
\hline
If things continue as they are, we will soon experience a major ecological disaster. & Attitude \\
\hline
In my daily life, I pay attention to how different types of waste are managed. & Behavior \\
\hline
I try to reduce the use of single-use items, such as bags and bottles. & Behavior \\
\hline
I prefer using public transportation or walking over driving a car. & Behavior \\
\hline
At home or in the office, I pay attention to saving electricity. & Behavior \\
\hline
When shopping, I tend to choose products that are more friendly to the environment, even if they might be more expensive. & Behavior \\
\hline
I use reusable items instead of single-use ones. & Behavior \\
\hline
\caption{Post-Survey Questions and Their Types.} \label{tab:post:survey} \\
\end{longtable}

% \newpage
% \twocolumn

\end{document}